\documentclass[aps,pra,twocolumn,amsmath,amssymb,nofootinbib,superscriptaddress,floatfix]{revtex4-1}
\allowdisplaybreaks
\usepackage{amsmath}
\usepackage{amssymb}
\usepackage{amsthm}
\usepackage[pdftex,colorlinks,citecolor=blue,linkcolor=red,urlcolor=blue]{hyperref} 
\usepackage{graphicx}
\usepackage{dcolumn} 
\usepackage{bm} 
\usepackage{longtable}
\usepackage{ulem}   
\usepackage[paperwidth=8.5in,paperheight=11in,centering,hmargin=2cm,vmargin=2.5cm]{geometry} 

\usepackage{comment} 
\usepackage{datetime}

\newcommand{\pri}[1]{\ensuremath #1^{\prime}}

\renewcommand\({\ensuremath \left(}
\renewcommand\){\ensuremath \right)}

\renewcommand\]{\ensuremath \right]}

\newcommand{\ket}[1]{{|\,{#1}\,\rangle}}
\newcommand{\bra}[1]{{\langle\,{#1}\,|}}

\newcommand{\ua}{{\uparrow}}

\def\beq{\begin{equation}}
\def\eeq{\end{equation}}
\def\bea{\begin{eqnarray}}
\def\eea{\end{eqnarray}}

\begin{document}

\title{ 
Stroboscopic Symmetry-Protected Topological Phases
      }

\author{Thomas Iadecola} 
\affiliation
{
Physics Department, Boston University, Boston, Massachusetts 02215, USA
} 

\author{Luiz H. Santos} 
\affiliation
{
Perimeter Institute for Theoretical Physics,
Waterloo, Ontario, N2L 2Y5, Canada
} 

\author{Claudio Chamon} 
\affiliation
{
Physics Department, Boston University, Boston, Massachusetts 02215, USA
} 

\date{\today}

\begin{abstract}
Symmetry-protected topological (SPT) phases of matter have been the focus of many recent theoretical investigations, but controlled mechanisms for engineering them have so far been elusive.  In this work, we demonstrate that by driving interacting spin systems periodically in time and tuning the available parameters, one can realize lattice models for bosonic SPT phases in the limit where the driving frequency is large.  We provide concrete examples of this construction in one and two dimensions, and discuss signatures of these phases in stroboscopic measurements of local observables.
\end{abstract}

\maketitle

\section{Introduction}
Since
the discovery of the quantum Hall effect (QHE)~\cite{QHE-Book}, topological phenomena in quantum many-body systems have dramatically changed our understanding of
phases of matter.
In particular, the study of the fractional QHE brought about the notion of topological order~\cite{Wen1990a,Wen1990b,Wen1990c}, which characterizes phases of matter with emergent fractional excitations and topological ground-state degeneracy, which cannot be described within the standard Landau-Ginzburg framework.

In recent years, the prediction and discovery of topological band
insulators~\cite{Hasan2010,Qi2011} has awakened a great deal of interest in gapped \textit{symmetry-protected} topological (SPT) phases of matter. 
These phases of matter lack fractionalized degrees of freedom, but display topological properties that manifest themselves in non-trivial boundary states that are protected by global symmetries. While they do not display the long-range entanglement of topologically-ordered systems, SPT phases of matter are characterized primarily by a nontrivial short-range entanglement structure in the low-energy states \cite{Chen-2012-2013}.

Following the classification of weakly-interacting
fermionic SPT states~\cite{Schnyder2008,Kitaev2009,Ryu2010},
there has been a vast amount of recent effort 
to classify strongly-interacting SPT 
phases~\cite{Chen-2012-2013,Lu-2012,Vishwanath-2013,Metliski-2013-a,C-Wang-2014,Bi-2013} as well as to construct models supporting them~\cite{Chen-2012-2013,Chen-2011,Levin-2012,Chen-2014,Burnell-2013,Geraedts-2014,Santos2015,Senthil2012,Regnault2013,Furukawa2013,Liu2014}. In light of this effort, it is highly desirable to identify controlled mechanisms capable of bringing SPT states into realization.  

In this paper, we put forward a proposal to realize bosonic SPT phases
as out-of-equilibrium states of quantum spin systems with periodically-driven multispin interactions.
The systems we study are described by time-dependent Hamiltonians of the form
\begin{equation}
\label{eq: H(t) in the lab frame}
\begin{split}
H(t)
=
H_{0}
+
\Theta(t)\,f(t)
\,
H_{\rm{int}}
\,,
\end{split}•
\end{equation}
where
$
H_{0}
$
is a local Hamiltonian describing a trivial paramagnet (i.e., one whose ground state is a trivial product state) and
$
H_{\rm{int}}
$
is a local interaction with a time-periodic coupling constant
$
f(t)
=
f(t+T)
$
with zero mean and a characteristic frequency $\omega = 2\,\pi/T$.
$\Theta(t)$ is the Heaviside function denoting a protocol 
where the drive is switched on at $t=0$.  

When $H_{\rm int}=0$, $H(t)=H_0$ can be mapped from a trivial paramagnetic Hamiltonian to an SPT Hamiltonian by a product of local unitary transformations that entangles the local degrees of freedom in a nontrivial way~\cite{Chen-2012-2013,Santos2015}.  Such transformations arise naturally in the study of many-body systems with periodically-driven interactions.  In particular, we will show that, in the limit of large $\omega$, the time-periodic unitary transformation to the ``rotating frame," (we set $\hbar=1$)
\begin{equation}
\label{eq: unitary transformation to rotating frame}
U_{\rm{R}}(t)
=
e^
{
\mathrm{i}\,
\int^{t}_{0}\,\mathrm{d}\,t'\,f(t')\,
H_{\rm{int}}
}
\equiv
e^
{
\mathrm{i}\,
g(t)\,
H_{\rm{int}}
}
\,,
\end{equation}
generates the desired entanglement if $H_{\rm int}$ is chosen appropriately.  The transformation $U_R(t)$ maps a state
$
\ket{\psi(t)}
$, whose time evolution is governed 
by the Hamiltonian~(\ref{eq: H(t) in the lab frame}),
into a state 
$
\ket
{
\psi_{\rm{R}}(t)
}
=
U_{\rm{R}}(t)
\,
\ket
{
\psi(t)
}
$
whose time evolution is generated by
\begin{align}\label{eq: H(t) in the rotating frame}
H_{\rm{R}}(t)&=U_{\rm{R}}(t)\,H(t)\,U^{\dagger}_{\rm{R}}(t)-\mathrm i\, U_{\rm{R}}(t)\,\partial_t\,U^{\dagger}_{\rm{R}}(t).
\end{align}
The stroboscopic
evolution of the initial state in the rotating frame,
$
\ket{\psi_{R}(nT)}
=
e^
{
-\mathrm{i}
\,
\mathcal{H}_{\rm F}\,
nT
}
\,
\ket{\psi_{R}(0)}
$
($n\in\mathbb Z$),
is governed by the Floquet Hamiltonian
$
\mathcal{H}_{\rm F}
$,
which can be systematically determined via a Magnus expansion~\cite{Blanes2009,Bukov2014}.
(Note that $\mathcal H_{\rm F}$ is also the generator of stroboscopic evolution in the ``lab frame," although we work with states in the rotating frame for convenience.)
In the infinite-frequency limit, the Floquet Hamiltonian is nothing but the 
time-average of $H_{\rm R}(t)$,
\begin{equation}
\label{eq: H Floquet infinite frequency limit}
\begin{split}
\mathcal{H}^{(0)}_{\rm F}
&=
\frac{1}{T}
\,
\int_{0}^{T}
\,
\mathrm{d}t\, 
H_{\rm R}(t)
\,,
\end{split}•
\end{equation}
while the $n$-th order term in the Magnus expansion is of order $1/\omega^n$.  We will refer to $\mathcal{H}^{(0)}_{\rm F}$ as the stroboscopic Hamiltonian, because in the infinite-frequency limit, where only $\mathcal{H}^{(0)}_{\rm F}$ survives, the stroboscopic evolution and the true unitary evolution of the time-dependent system coincide.
\begin{figure}
\centering
\includegraphics[width=.45\textwidth]{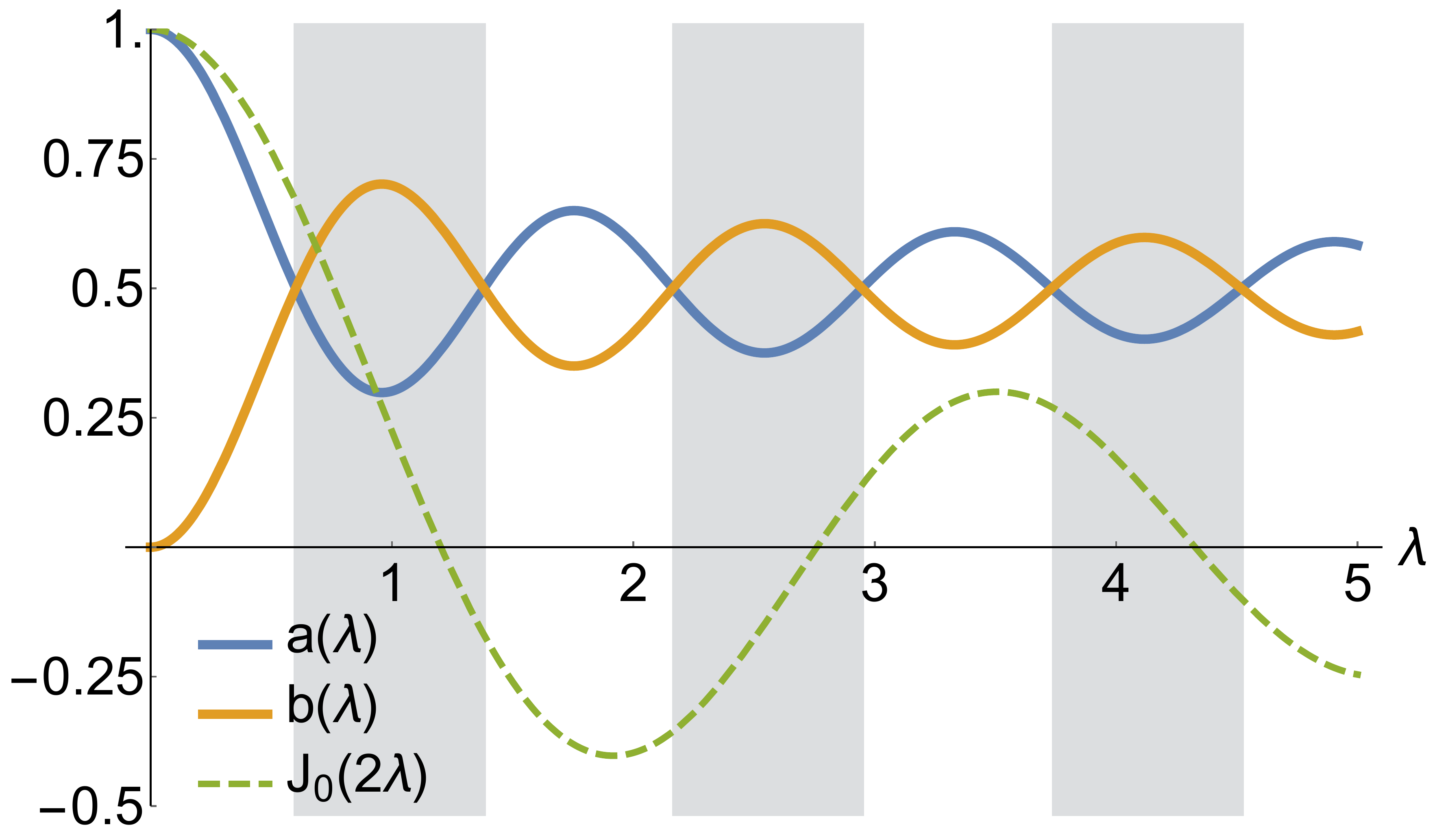}
\caption{(Color online) Couplings in the leading term of the 
Magnus expansion~(\ref{eq: H Floquet infinite frequency limit}) as functions of the scaled driving amplitude $\lambda$. White and gray regions correspond,
respectively, to trivial and stroboscopic SPT phases.}\label{fig: couplings}
\end{figure}

If the amplitude of the drive is small compared to the frequency, the stroboscopic Hamiltonian~(\ref{eq: H Floquet infinite frequency limit}) 
simply reduces to $H_{0}$.
On the other hand, when the amplitude of the drive is chosen to scale with the frequency $\omega$, the stroboscopic Hamiltonian can acquire a nontrivial form that is different from $H_0$~\cite{Bukov2014}. In this work, we show that the stroboscopic Hamiltonian~(\ref{eq: H Floquet infinite frequency limit})
describes microscopic models of SPT states with $\mathbb{Z}_{2}\times\mathbb{Z}_{2}$~\cite{Chen-2012-2013,Geraedts-2014,Santos2015}
and $\mathbb{Z}_{2}$~\cite{Chen-2012-2013,Levin-2012}  symmetries, respectively, for one- and two-dimensional driven systems.  We refer to the phases generated in this way as \textit{stroboscopic} SPT (SSPT) phases.
Remarkably, we find that, while the SSPT Hamiltonian~(\ref{eq: H Floquet infinite frequency limit})
is invariant under the global symmetry, the original time-dependent
Hamiltonian~(\ref{eq: H(t) in the lab frame}) is not.
Hence the global symmetry of the SSPT phase is found to be an emergent property of the high-frequency limit of $\mathcal H_{\rm F}$.
These results can be generalized to other symmetry classes.  
Finally, we also demonstrate that the dynamics of local observables at stroboscopic times can be used to probe the nontrivial edge states of SSPT systems without the need to prepare the system in the ground state of $\mathcal H_{\rm F}$.


\section{$\mathbb Z_2\times\mathbb Z_2$ SSPT Phase in 1D}
\subsection{The model}
We begin by studying an open 1D chain with $N$ sites described by the time-dependent Hamiltonian \eqref{eq: H(t) in the lab frame} with
\begin{equation}
\label{eq: 1D Hamiltonian in the lab frame}
\begin{split}
&\,
H_{1\rm{D}}(t)
=
h\,
\sum^{N}_{i=1}\,
\sigma^{x}_{i}
+
\Theta(t)\,f(t)\,
\sum^{N-1}_{i=1}\,
\sigma^z_{i}\,
\sigma^z_{i+1}
\,,
\end{split}
\end{equation}
and $f(t)=\lambda\,\omega\, \cos(\omega t + \varphi)$ ($\lambda > 0$). Note that the driving amplitude is taken to scale linearly with the frequency, so that $\lambda$ is dimensionless.
The Pauli operators $\sigma^a_{i}$ ($a=x,y,z$) satisfy the onsite algebra
$
[\sigma^a_{i},\sigma^b_j]
=
2 \mathrm i\,\delta_{ij}\,  \epsilon_{abc}\, \sigma^c_i
$
and the anticommutation relation
$
\{\sigma^a_{i},\sigma^b_i\}=2\, \delta_{ab}\, 
$.
Furthermore, notice that the Hamiltonian~(\ref{eq: 1D Hamiltonian in the lab frame})
has an onsite $\mathbb Z^{\,}_{2}$ spin flip symmetry generated by 
$
S=\prod^{N}_{i=1}\sigma^x_{i}
$.
Henceforth, we set the energy scale $h=1$, with the understanding that the limit $\omega\to\infty$ corresponds to taking $\omega \gg h$.

Upon making the transformation to the rotating frame, we find that, for $\varphi=0$, the stroboscopic 
Hamiltonian~(\ref{eq: H Floquet infinite frequency limit}) is given by
\begin{equation}
\label{eq: finite chain}
\begin{split}
\mathcal{H}_{\mathbb Z_2\times\mathbb Z_2}
&\,=
J_{0}(2\lambda)\, (\sigma^x_1+\sigma^x_N)
\\
&\,
+
\sum_{i=2}^{N-1}
\,
\[a(\lambda)\,
\sigma^{x}_{i}
-
b(\lambda)\,
\sigma^{z}_{i-1}\sigma^{x}_{i}\sigma^{z}_{i+1}\]
\,,
\end{split}
\end{equation}
where
$
a(\lambda)
=
\frac{1}{2}
\left[
1
+
J_{0}(4\lambda)
\right]
$,
$
b(\lambda)
=
1
-
a(\lambda)
$
and
$
J_{0}(x)
$
is the Bessel function of the first kind.

Observe that the Hamiltonian~(\ref{eq: finite chain})
possesses a global $\mathbb{Z}_{2}\times\mathbb{Z}_{2}$
symmetry generated by
$
S_{\rm even}
=
\prod_{i\ \mathrm{even}}
\sigma^{x}_{i}
$
and
$
S_{\rm odd}
=
\prod_{i\ \mathrm{odd}}
\sigma^{x}_{i},
$
corresponding to independent spin flips on the even and odd sublattices.
However, the time-dependent 
Hamiltonian~(\ref{eq: 1D Hamiltonian in the lab frame})
has a $\mathbb Z_2$ symmetry, rather than a $\mathbb Z_2\times\mathbb Z_2$ symmetry -- in other words, this enlarged symmetry group is an emergent property of the high-frequency limit $\omega\to\infty$, as it appears only upon taking the 
time average Eq.~(\ref{eq: H Floquet infinite frequency limit}).

We plot the couplings $a(\lambda)$, $b(\lambda)$, and $J_{0}(2\lambda)$ in Fig.~\ref{fig: couplings}.  By varying $\lambda$, one can tune the couplings such that $a(\lambda)>b(\lambda)$ or vice versa.  We will argue below that the values of $\lambda$ for which $a(\lambda)=b(\lambda)$ are critical points of the effective Hamiltonian that separate a trivial insulating phase from an SPT phase.

\begin{figure}[b]
\centering
\includegraphics[width=.475\textwidth]{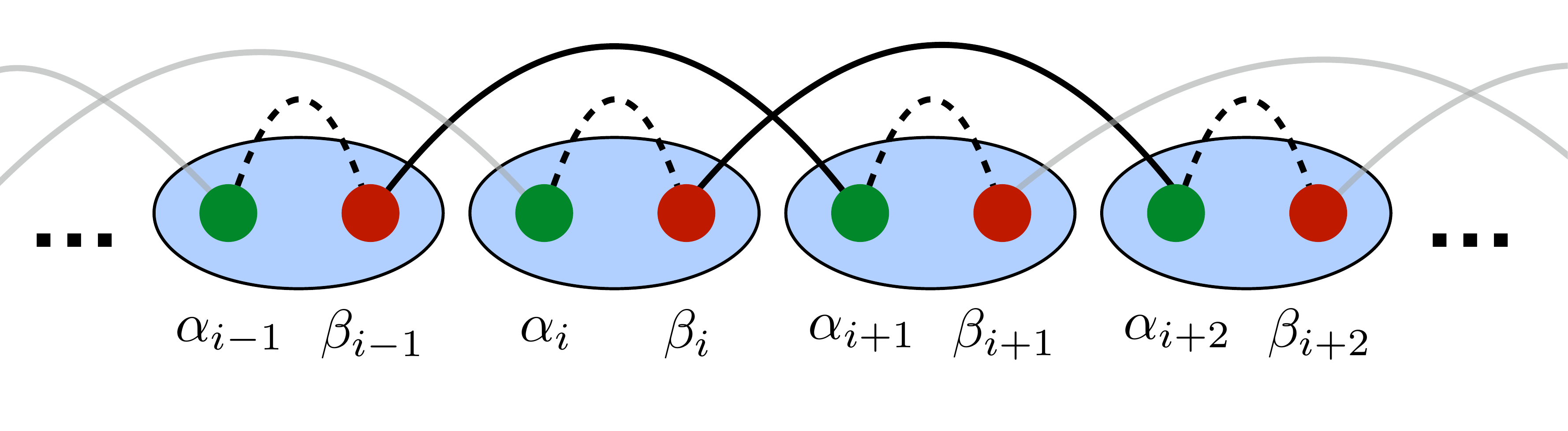}
\caption{(Color online) Competing dimerization patterns in the $\mathbb Z_2\times\mathbb Z_2$ SPT chain. 
Dotted and solid lines account, respectively, for the dominant dimerization
patterns in the trivial and SPT phases. When only the $3$-spin term is present in the model, a dangling spin is localized on the edges.}\label{fig: dimerization}
\end{figure}

To continue our analysis of this $\mathbb Z_2\times\mathbb Z_2$-symmetric Hamiltonian, it is instructive to rewrite it in terms of Majorana operators $\alpha_i = (\prod_{j<i} \sigma^x_j)\sigma^z_i$ and $\beta_i =\mathrm i \, \alpha_i\, \sigma^x_i$,~\cite{Kitaev2001,Fendley2012} which are Hermitian and satisfy the usual fermionic algebra.  In terms of these operators, the Hamiltonian~\eqref{eq: finite chain} reads
\begin{align}\label{majorana hamiltonian}
\begin{split}
\mathcal{H}_{\mathbb Z_2\times\mathbb Z_2}&=-\mathrm i\, J_0(2\lambda)(\alpha_1\, \beta_1+\alpha_N\, \beta_N)\\
&\indent-\mathrm i\, a(\lambda)\sum^{N-1}_{i=2}\alpha_i\, \beta_i +\mathrm i\, b(\lambda)\sum^{N-1}_{i=2}\beta_{i-1}\, \alpha_{i+1}.
\end{split}
\end{align}
This Hamiltonian contains two types of terms that can be thought of as projectors onto two distinct dimerization patterns that encode the entanglement structure of the ground-state wavefunction (see Fig.~\ref{fig: dimerization}).  The pattern encoded by the $\alpha_i\beta_i$ terms involves Majorana dimers on each site.  It is ``trivial" in the sense that, for a finite chain, the pattern pairs all Majorana operators.  On the other hand, the $\beta_{i-1}\alpha_{i+1}$ terms encode dimerization between next-neighbor Majoranas of opposite types.  This pattern is ``nontrivial" in the sense that it leaves two unpaired Majoranas at each end of a finite chain, yielding a fourfold ground-state degeneracy as a signature of the entanglement structure of the SPT phase.  Equivalently, one can see this fourfold degeneracy from Eq.~\eqref{eq: finite chain}, as the operators $\sigma^z_1$ and $\sigma^z_N$ commute with the Hamiltonian if only the three-spin interaction contributes.

The transition between these two patterns and the associated phases occurs at the point $a(\lambda)=b(\lambda)$, where the bulk gap closes.  To see this, we combine the Majorana operators into complex fermions $c^\dagger_{i}=(\alpha_i+\mathrm i\, \beta_{i})/2$, in terms of which the Hamiltonian becomes
\begin{align}\label{kitaev p-wave}
\begin{split}
\mathcal{H}_{\mathbb Z_2\times\mathbb Z_2}&=t\sum^{N-1}_{i=2}(c^\dagger_{i+1}c^{\,}_{i-1}+c^{\,}_{i+1}c^{\,}_{i-1}+\text{H.c.})\\
&\indent +\mu\sum^{N-1}_{i=2}c^\dagger_ic^{\,}_i+\mu_0(c^\dagger_1c^{\,}_1+c^\dagger_Nc^{\,}_N)
\end{split}
\end{align}
where we have dropped a constant term, and where we have defined $t =  b(\lambda)$, $\mu = 2\, a(\lambda)$, and $\mu_0=2J_0(2\lambda)$.  This model is nothing but two decoupled copies (one on the even sublattice and one on the odd) of the Kitaev model for a 1D $p$-wave superconductor~\cite{Kitaev2001}.  The critical point for this model is well-known, and occurs for $\mu=2t$.  However, this is equivalent to the condition $a(\lambda)=b(\lambda)$.  It is important to note that while the chemical potential $\mu_0$ at the ends of the chain is not equal to the bulk value $\mu$, the location of the transition is not affected for a sufficiently long chain, as we have verified by exact diagonalization.

The preceding discussion illustrates that the stroboscopic Hamiltonian \eqref{eq: finite chain} is $\mathbb Z_2\times\mathbb Z_2$-symmetric and contains one free parameter, $\lambda$, that tunes the system across the transition between the $\mathbb Z_2\times\mathbb Z_2$ SSPT phase and the trivial paramagnetic phase.  However, it is evident from Fig.~\ref{fig: couplings} that the coupling $a(\lambda)\neq 0$ for any $\lambda$. Naively, then, it seems that one cannot access the ``ideal" scenario where the operators $\sigma^z_{1,N}$ commute with the Hamiltonian.  Nevertheless, this is not the case, as the local field $J_0(2\lambda)$ at the ends of the chain vanishes identically if $2\lambda$ is equal to a zero of the Bessel function $J_0$ (see Fig.~\ref{fig: couplings}).  In this case, the operators $\sigma^z_1$ and $\sigma^z_N$ identically commute with the Hamiltonian \eqref{eq: finite chain}, and the system has an exact fourfold ground-state degeneracy despite the presence of a transverse field in the bulk.  Deviations from these special values of $\lambda$ split this degeneracy by an amount that decreases exponentially with system size, and the system remains in the SSPT phase so long as the bulk gap remains open.

\begin{figure*}
\centering
\hspace{-.5cm}(a)\hspace{.5cm}\includegraphics[width=.4\textwidth]{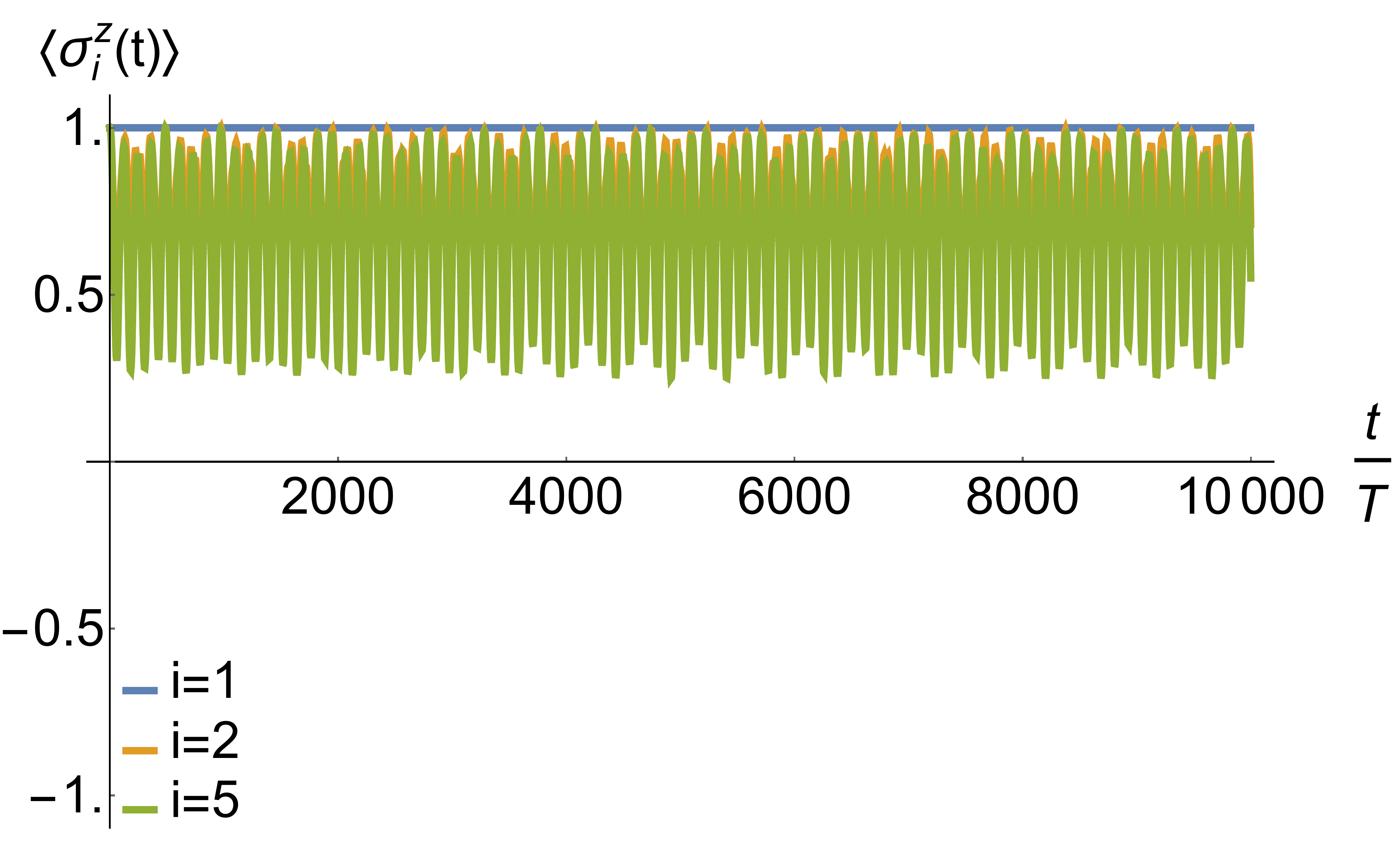}\qquad\qquad(d)\hspace{.5cm}\includegraphics[width=.4\textwidth]{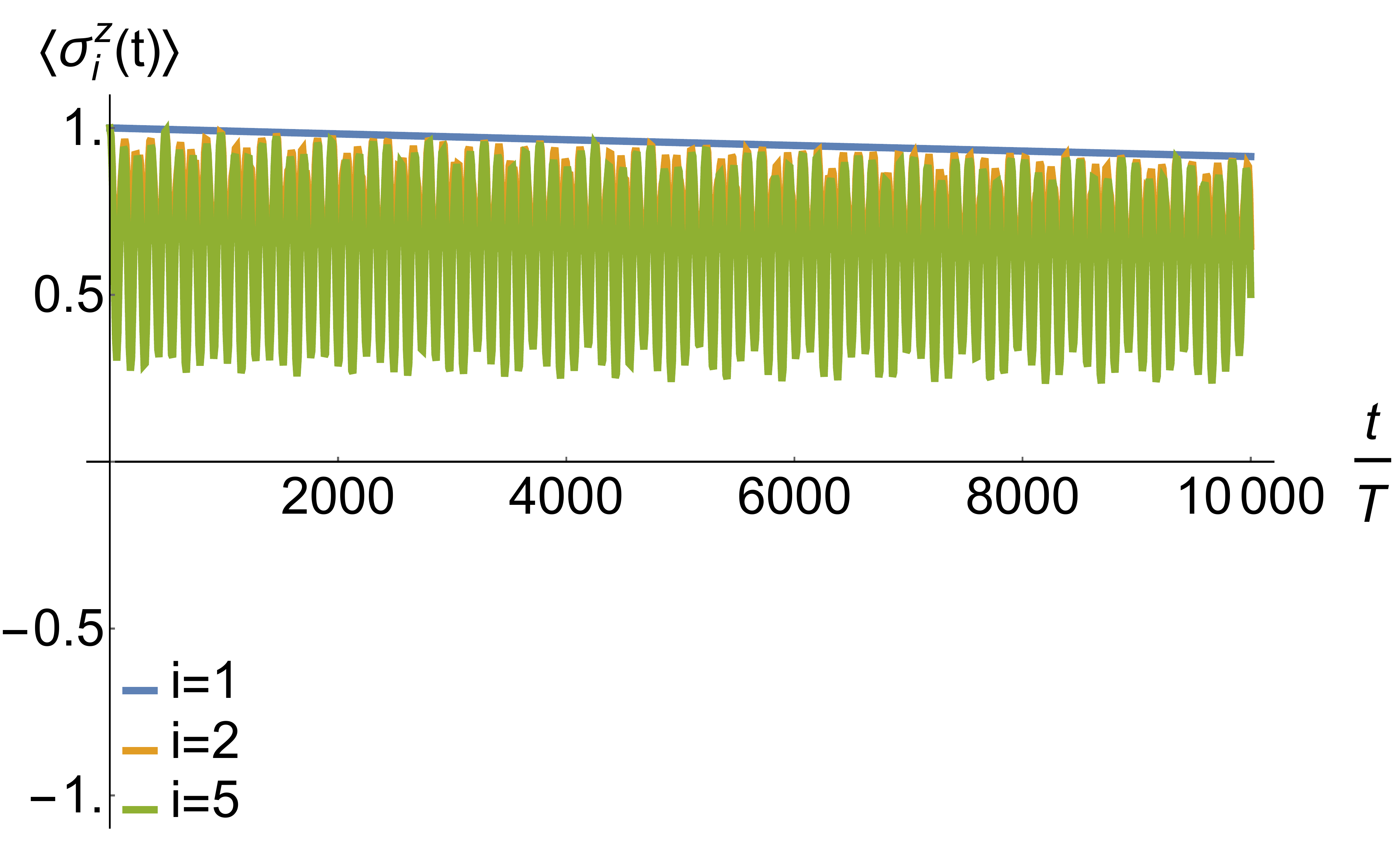}\\
\hspace{-.5cm}(b)\hspace{.5cm}\includegraphics[width=.4\textwidth]{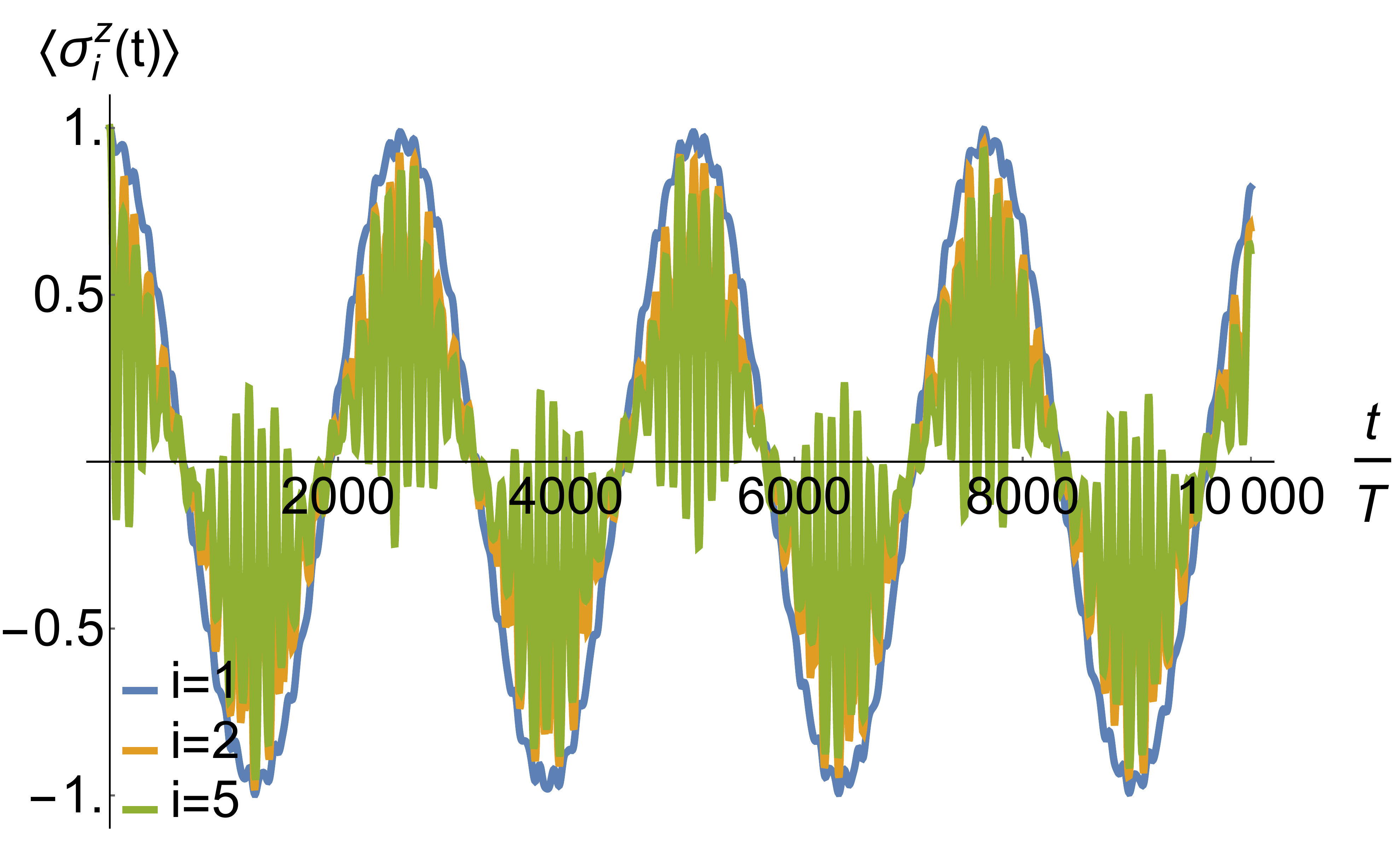}\qquad\qquad(e)\hspace{.5cm}\includegraphics[width=.4\textwidth]{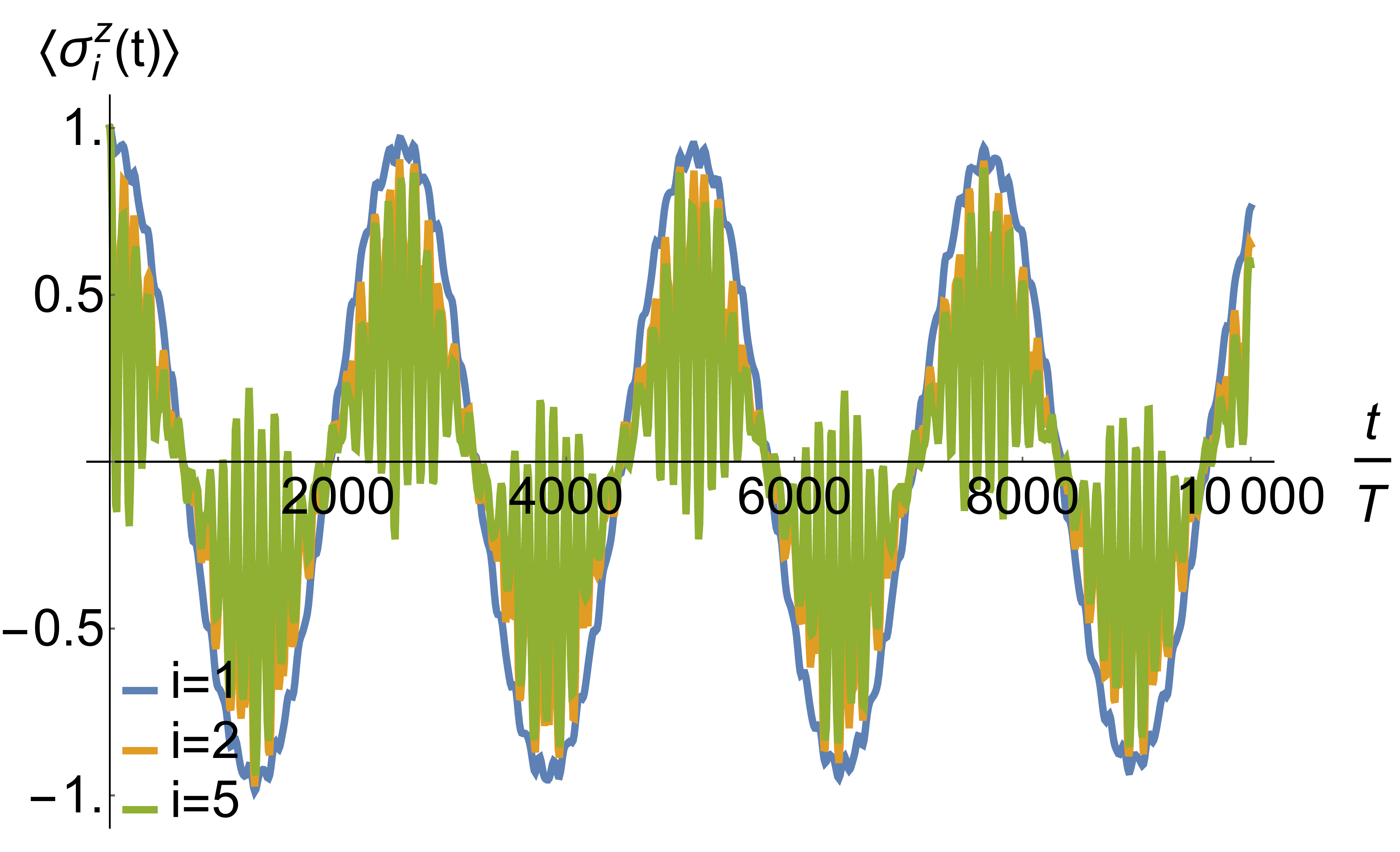}\\
\hspace{-.5cm}(c)\hspace{.5cm}\includegraphics[width=.4\textwidth]{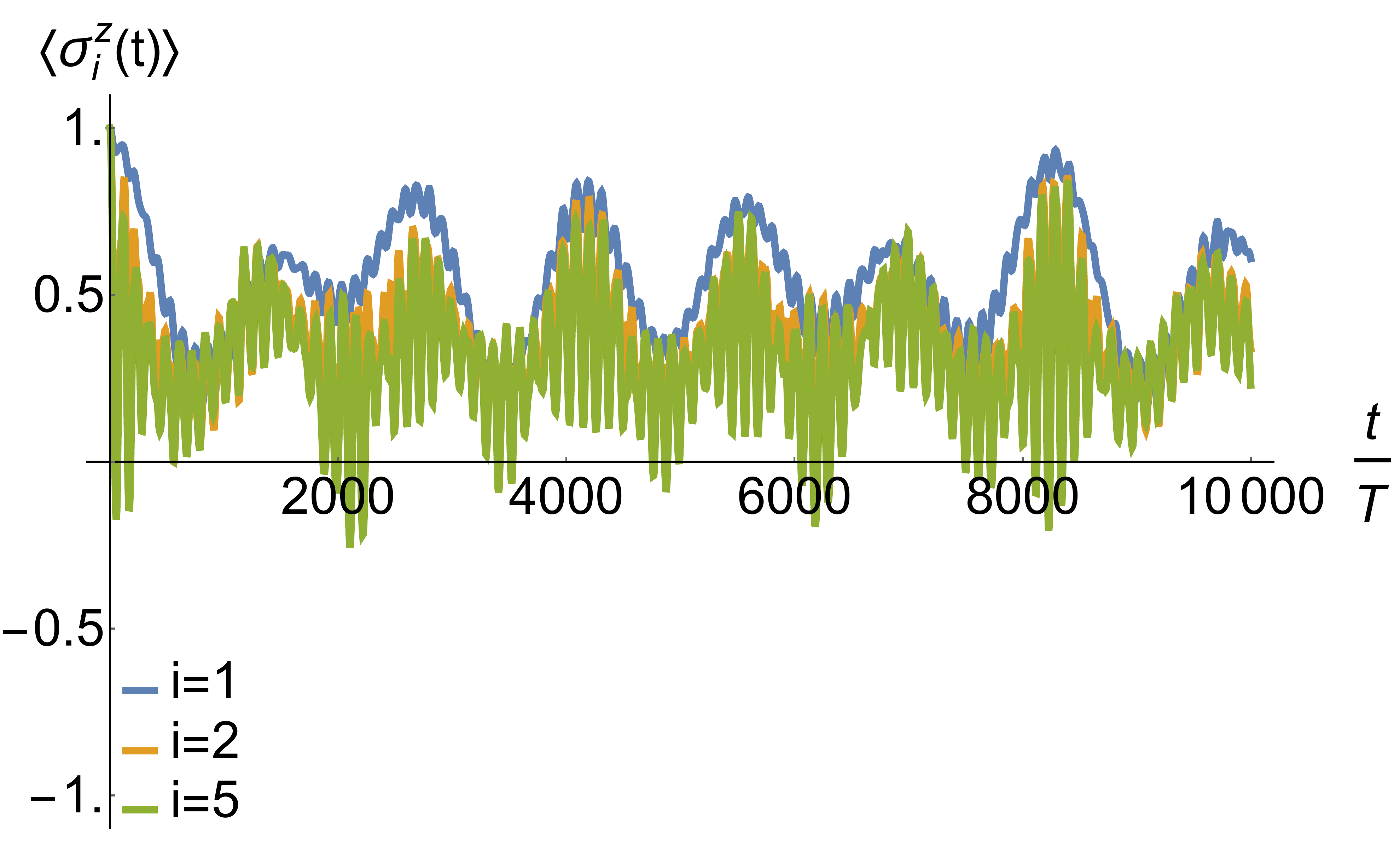}\qquad\qquad(f)\hspace{.5cm}\includegraphics[width=.4\textwidth]{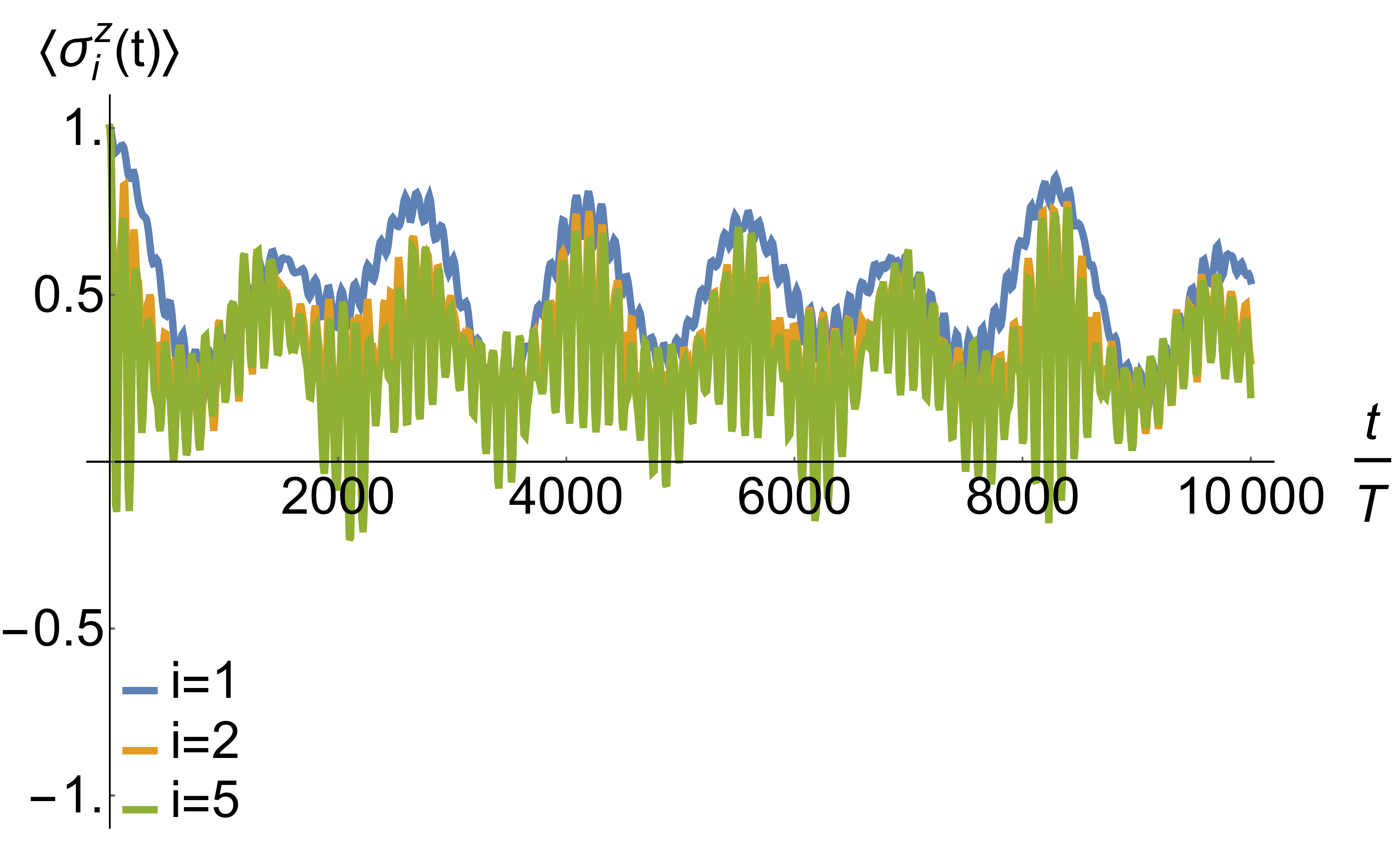}\\
\caption{(Color online) Stroboscopic evolution of $\langle\sigma^z_i(t)\rangle$ for $i=1,2,$ and $5$ for an eight-site chain, where the initial state is chosen to be the product state $\ket{\Psi_0}=\ket{\ua\ua\dots\ua}$.  Panels (a)--(c) depict stroboscopic evolution at $\omega=\infty$ [\textit{i.e.}~as defined in Eq.~\eqref{stroboscopic dynamics}], while panels (d)--(f) depict the exact stroboscopic evolution [\textit{i.e.}~as defined in Eq.~\eqref{stroboscopic dynamics finite frequency}] for $\omega = 100\, h$.  In (a), $2\lambda$ equals the second zero of the Bessel function $J_{0}(x)$ in Fig.~\ref{fig: couplings}, and the longitudinal symmetry-breaking field $h_z=0$.  In (b), $\lambda = 2.6$ and $h_z=0$, while in (c), $\lambda=2.6$ and $h_z=0.01$ in units of the bare transverse field $h$.  The plots in panels (d), (e), and (f) use the same parameters as the ones in panels (a), (b), and (c), respectively.
}\label{fig: sigmaZ}
\end{figure*}

\subsection{Signatures in stroboscopic dynamics}
\subsubsection{Infinite-frequency limit}
So far, we have demonstrated that, in the limit $\omega\to\infty$, the stroboscopic evolution of the periodically-driven spin chain of Eq.~\eqref{eq: 1D Hamiltonian in the lab frame} is generated by the effective Hamiltonian~\eqref{eq: finite chain}, for an appropriately-chosen driving protocol.  However, it remains to be shown that the stroboscopic evolution generated by Eq.~\eqref{eq: finite chain} yields telltale signatures of SPT physics in local measurements.  To address this point, we consider the local spin expectation value
\begin{align}\label{stroboscopic dynamics}
\langle\sigma^z_i(t)\rangle&=\bra{\Psi_0}e^{+\mathrm i\, \mathcal{H}_{\mathbb Z_2\times\mathbb Z_2}  t}\, \sigma^z_i\, e^{-\mathrm i\, \mathcal{H}_{\mathbb Z_2\times\mathbb Z_2}  t}\ket{\Psi_0},
\end{align}
where $\ket{\Psi_0}$ is some initial state.  This quantity coincides with the true time evolution of the operator $\sigma^z_i$ in the limit $\omega\to\infty$, where the period $T$ is infinitesimally small and $t=nT$ ($n\in\mathbb Z$) is approximately a continuous variable.  For simplicity, we choose the initial state $\ket{\Psi_0}$ to be a product of eigenstates of $\sigma^z_j$ on each site $j$, so that $\langle\sigma^z_i(t)\rangle$ is invariant under  $U_{\rm R}(t)$ (\textit{i.e.,} the unitary transformation to the rotating frame).

The observable defined in Eq.~\eqref{stroboscopic dynamics} provides a clear signature of the edge states in the SSPT phase, even though the product state $\ket{\Psi_0}$ is a highly out-of-equilibrium state with respect to the stroboscopic Hamiltonian (c.f.~Ref.~\cite{vishwanath}).  When $J_0(2\lambda)=0$, $\langle\sigma^z_1(t)\rangle$ and $\langle\sigma^z_N(t)\rangle$ are independent of time, since $\sigma^z_1$ and $\sigma^z_N$ commute with the effective Hamiltonian $\mathcal{H}_{\mathbb Z_2\times\mathbb Z_2}$.  For $i\neq 1$ or $N$, however, $\langle\sigma^z_i(t)\rangle$ evolves quasi-periodically in time, with oscillations occurring on a timescale $\tau_{\rm c}$ on the order of the inverse bulk energy gap of $\mathcal{H}_{\mathbb Z_2\times\mathbb Z_2}$ [see Fig.~\ref{fig: sigmaZ}(a)]. If $\lambda$ is tuned slightly away from one of these special values but remains within the phase boundary, which amounts to adding a small transverse field $(\sigma^x_1+\sigma^x_N)$, then the end spins precess in the $y$-$z$ plane on timescales much longer than $\tau_{\rm c}$, so that the bulk and boundary behavior can be distinguished. Because the end spins were completely unconstrained before the addition of the transverse field, this precession, which is characteristic of a free spin placed in a magnetic field perpendicular to the quantization axis, manifests itself in oscillations of $\langle\sigma^z_{1,N}(t)\rangle$ about zero [see Fig.~\ref{fig: sigmaZ}(b)].  In this way, $\langle\sigma^z_{1,N}(t)\rangle$ can be used to distinguish the free edge spins characteristic of SPT physics from spins that are ``frozen" due to the presence of a symmetry-breaking field. For example, if a small longitudinal field $h_{z}\sum^N_{i=1}\sigma^z_i$ is added to Eq.~\eqref{eq: 1D Hamiltonian in the lab frame}, thereby breaking the $\mathbb Z_2\times\mathbb Z_2$ symmetry of the stroboscopic Hamiltonian, then one finds that the end spins no longer wrap the unit circle in the $y$-$z$ plane as they precess when the transverse field is added at the ends of the chain.  Consequently, $\langle\sigma^z_{1,N}(t)\rangle$ no longer oscillates around zero, but around a nonzero value whose sign matches the sign of the longitudinal field [see Fig.~\ref{fig: sigmaZ}(c)].  Thus, in the infinite-frequency limit, stroboscopic measurements of $\langle\sigma^z_{i}(t)\rangle$ are an effective dynamical probe of SPT physics at the boundary of the driven system.

\subsubsection{Finite-frequency corrections}

The preceding discussion is an accurate description of the driven system in the limit where the driving frequency $\omega$ is infinite.  However, at any finite frequency, there are corrections to this behavior that become important in the infinite-\textit{time} limit, where the ``error" due to these corrections can accumulate without bound.  We now characterize the nature of these corrections (\textit{c.f.}~Ref.~\cite{Bukov2014}), and present arguments and numerical results showing that there is a window of time after the drive is switched on during which the signatures of the infinite-frequency SSPT phase can be observed at large but finite driving frequencies.

As the driving frequency $\omega$ is decreased, two effects occur that lead to deviations from the infinite-frequency case discussed in the previous section.  First, the stroboscopic evolution of observables, as in Eq.~\eqref{stroboscopic dynamics}, no longer coincides with the true time evolution of the system.  In particular, expectation values of observables become dressed by intra-period effects that become significant if the system is not observed at stroboscopic times $t_{n}=nT$ for $n\in\mathbb Z$~\cite{Bukov2014}.  However, the expectation values of observables at stroboscopic times are still predicted by the unitary evolution generated by the Floquet Hamiltonian $\mathcal H_{\rm F}$.  For instance, the stroboscopic evolution of $\sigma^z_i$, which is given at infinite driving frequency by Eq.~\eqref{stroboscopic dynamics}, becomes
\begin{subequations}
\begin{align}\label{stroboscopic dynamics finite frequency}
\langle\sigma^z_i(t_n)\rangle&=\bra{\Psi_0}e^{+\mathrm i\, \mathcal{H}_{\rm F}  t_n}\, \sigma^z_i\, e^{-\mathrm i\, \mathcal{H}_{\rm F}  t_n}\ket{\Psi_0},
\end{align}
where
\begin{align}
\mathcal H_{\rm F}&=\mathcal H^{(0)}_{\rm F}+\mathcal H^{(1)}_{\rm F}+\dots
\end{align}
contains all finite-frequency corrections $\mathcal H^{(k)}_{\rm F}\sim h\,  (h/\omega)^k$ to the infinite-frequency Floquet Hamiltonian $\mathcal H^{(0)}_{\rm F}$.
\end{subequations}

This brings us to the second effect, namely the fact that $\mathcal H_{\rm F}$ acquires finite-frequency corrections appearing at orders $1/\omega$ and higher in the Magnus expansion.  These corrections generically break the symmetry that protects the SSPT phase.  Indeed, in Appendix A, we present the leading finite-frequency correction to the infinite-frequency Hamiltonian $\mathcal H_{\mathbb Z_2\times\mathbb Z_2}$ that break the emergent $\mathbb Z_2\times\mathbb Z_2$ symmetry down to the $\mathbb Z_2$ symmetry of the original time-dependent Hamiltonian \eqref{eq: 1D Hamiltonian in the lab frame}.  We will now argue that these symmetry-breaking corrections are unimportant for the detection of the stroboscopic signatures of the SSPT phase discussed in the previous section, provided that the driving frequency $\omega$ is sufficiently large compared to the bare transverse field energy scale $h$.

To see this, observe that at some stroboscopic time $t_n=n\, T$, the $k$-th order Magnus correction is only important if $\mathcal H^{(k)}_{\rm F}\, t_n$ is comparable in size (modulo $2\pi$) to a number of order one.  Since $H^{(k)}_{\rm F}\sim h\, (h/\omega)^k$, this means that one must wait a (stroboscopic) time 
\begin{align}
n_*\equiv\frac{t_*}{T} \sim \(\frac{\omega}{h}\)^{k+1}
\end{align}
in order for symmetry-breaking effects to begin to manifest themselves.  Note that, in the limit $\omega\to\infty$, the time $t_*=n_*\, T\to\infty$ as well, so that the infinite-frequency limit manifests the enlarged $\mathbb Z_2\times\mathbb Z_2$ symmetry, as expected.

To support this argument in the context of the dynamical signatures of the SSPT phase discussed in the previous section, we have supplemented the infinite-frequency stroboscopic evolution [\textit{c.f.}~Eq.~\eqref{stroboscopic dynamics}] displayed in Fig.~\ref{fig: sigmaZ}(a)--(c) with finite-frequency calculations [\textit{c.f.}~Eq.~\eqref{stroboscopic dynamics finite frequency}] over a range of frequencies.  The finite-frequency stroboscopic calculations were performed using the exact evolution operator over a period, determined by direct numerical integration of the Schr\"odinger equation.  We find that a driving frequency $\omega \sim 100\, h$ is sufficiently large to extract the dynamical information necessary in order to infer the existence of symmetry-protected edge states in the manner outlined in the previous section [see Fig.~\ref{fig: sigmaZ}(d)--(f)].  Differences between these finite-frequency results and their infinite-frequency counterparts only begin to manifest themselves on timescales of order $t_* = 10^4\, T$, where, according to the scaling argument in the previous paragraph, the $k=1$ correction begins to become important.

\begin{figure}
\centering
(a)\includegraphics[width=.425\textwidth]{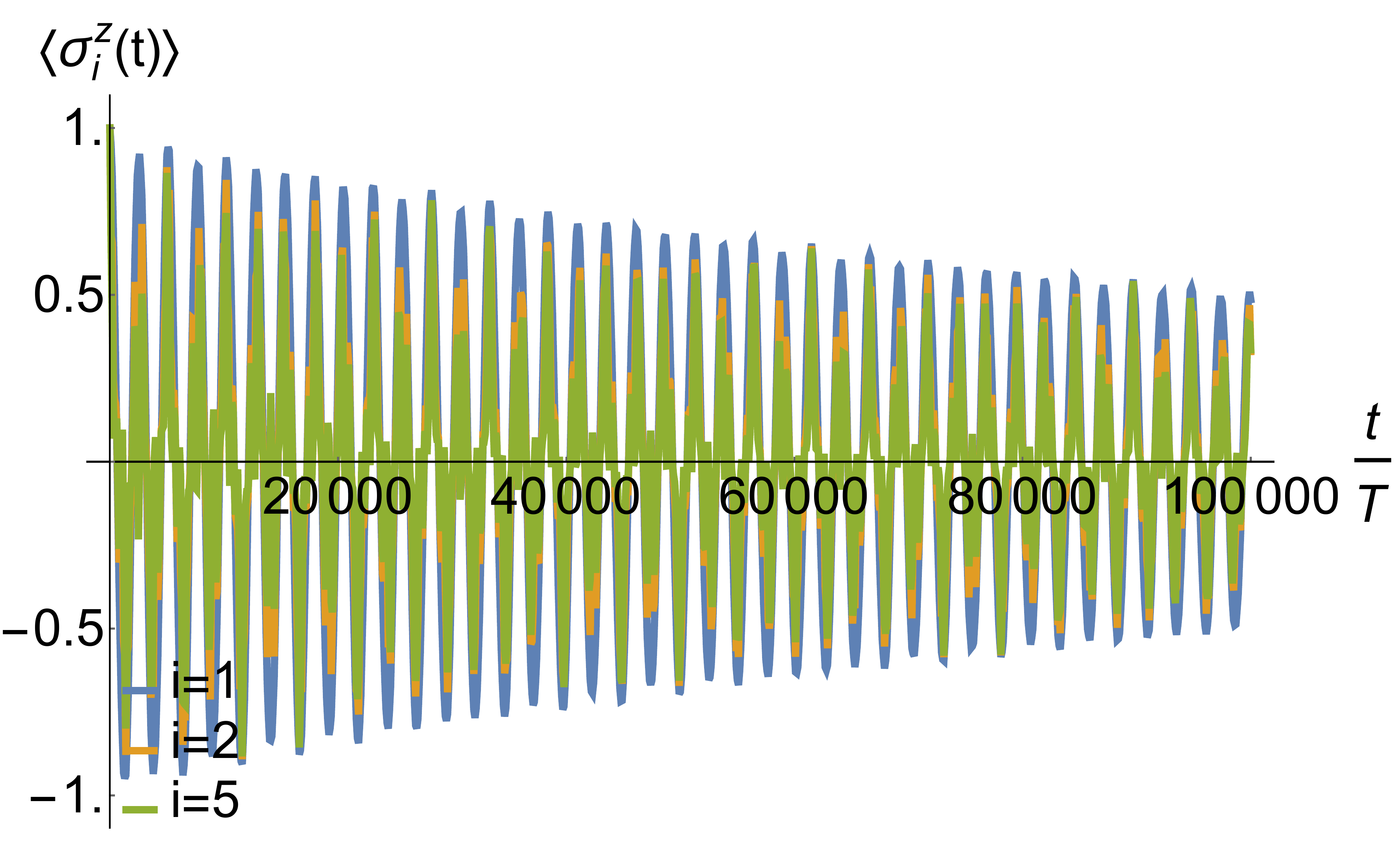}\\
(b)\includegraphics[width=.425\textwidth]{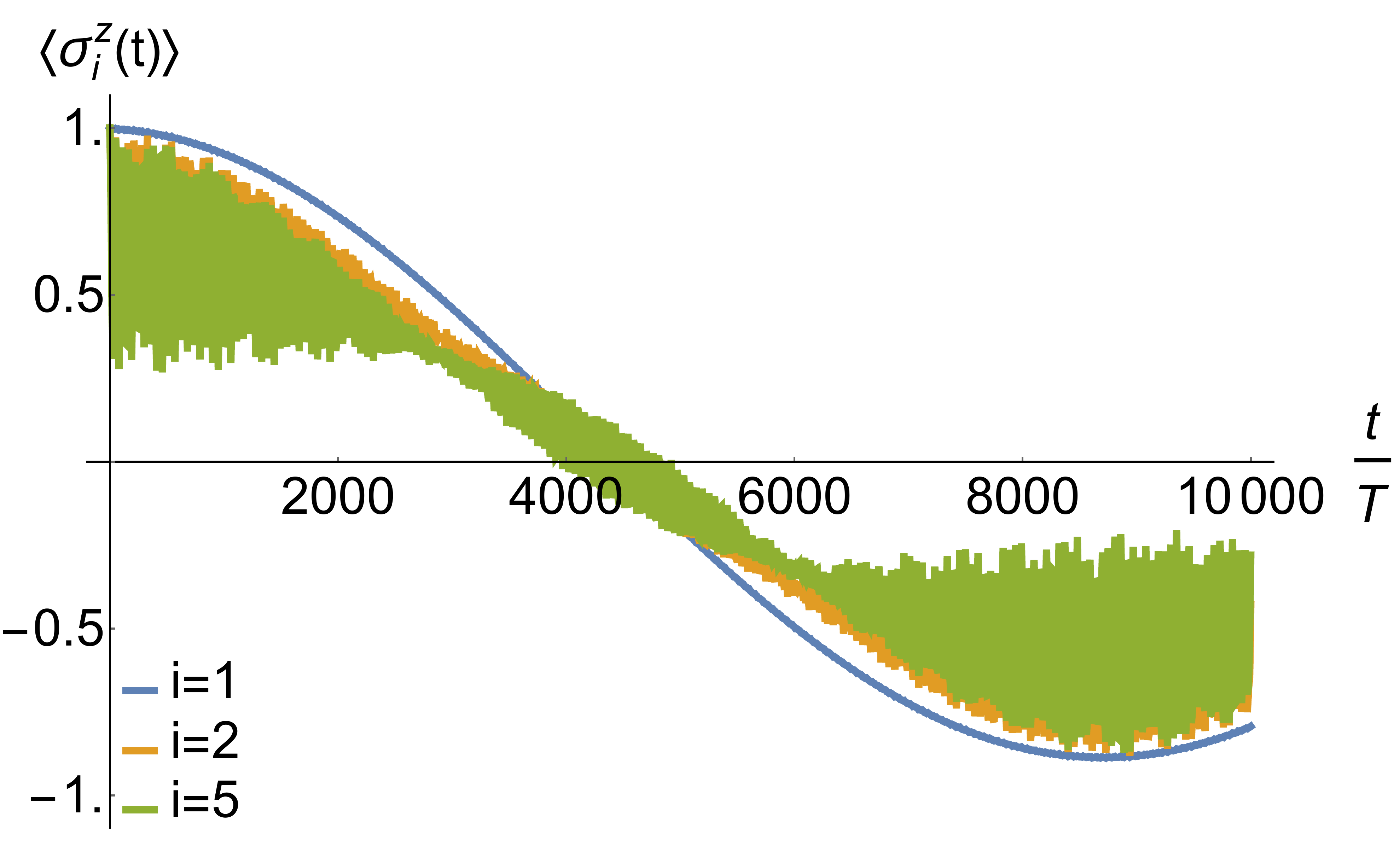}\\
\caption{(Color online) Finite-frequency corrections to stroboscopic evolution of $\langle\sigma^z_i(t)\rangle$.  The system size and initial state are the same as in Fig.~\ref{fig: sigmaZ}.  Panel (a) shows the evolution depicted in Fig.~\ref{fig: sigmaZ}(e) over a longer time, so that the envelope timescale discussed in the main text is more apparent.  Panel (b) uses the same parameters as Fig.~\ref{fig: sigmaZ}(a) and (d), but at frequency $\omega=10\, h$.
}\label{fig: exact_dynamics}
\end{figure}

The main difference that arises at large but finite frequencies is the appearance of an envelope timescale, much longer than $t_*$, on which the end spins oscillate (see Fig.~\ref{fig: exact_dynamics}).  For example, in Fig.~\ref{fig: exact_dynamics}(a), the dynamics shown in Fig.~\ref{fig: sigmaZ}(e) is plotted out to a time $t=10\, t_*$, where the appearance of the envelope timescale is very clear.  At the ``magic" values of $\lambda$ where the end spins are frozen in the infinite-frequency limit [\textit{c.f.}~Fig.~\ref{fig: sigmaZ}(a)], the envelope timescale also manifests itself in oscillations of the previously-frozen end spins due to the appearance of symmetry-breaking finite-frequency corrections at long times.  This is already evident in Fig.~\ref{fig: sigmaZ}(d), where $\omega=100\, h$, and is exaggerated in Fig.~\ref{fig: exact_dynamics}(b), where the frequency has been lowered to $10\, h$.

To summarize this discussion, we have argued in this section that the dynamical signatures of the SSPT phase discussed in the previous section are immune to finite-frequency corrections for a window of time whose size increases monotonically with the driving frequency and approaches infinity in the infinite-frequency limit.  At high but still reasonable frequencies $\omega\sim 100\, h$, this time window is sufficiently large to extract these dynamical signatures before the corrections begin to take over.

\section{Rationale, generalizations, and possible experiments}
The rationale behind the ability to engineer the SSPT Hamiltonian~(\ref{eq: finite chain}) can be stated as follows. 
First, recall that, when the driving vanishes, the mapping from the trivial phase to the SPT one can be achieved
via a product of local unitary transformations~\cite{Chen-2012-2013}. In the 
$\mathbb{Z}_{2}\times\mathbb{Z}_{2}$ case, the generator of this transformation is proportional to the Ising interaction in Eq.~(\ref{eq: 1D Hamiltonian in the lab frame})~\cite{Santos2015}.
On the other hand, in the driven system, the unitary transformation $U_{\rm{R}}(t)$ to 
the rotating frame is also generated by the Ising interaction.  Consequently, at infinite frequency, we found parameter regimes in which this transformation effectively
mapped a trivial paramagnet to an SPT one, and gave rise to an emergent 
$\mathbb{Z}_{2}\times\mathbb{Z}_{2}$ symmetry that is not shared by the time-dependent Hamiltonian~(\ref{eq: 1D Hamiltonian in the lab frame}).

The above discussion suggests a principle for obtaining an SSPT phase
in a periodically-driven system: the interaction term in Eq.~\eqref{eq: H(t) in the lab frame} should be chosen to be the generator
of the unitary transformation connecting a trivial to an SPT system.
In order to demonstrate that this stroboscopic approach to SPT phases applies 
beyond the 1D case discussed above, we consider a 
2D system on the triangular lattice with a driven three-spin interaction,
\begin{equation}
\label{eq: 2D driven Hamiltonian}
\begin{split}
H_{2\rm{D}}(t)
=
-h\,
\sum_{j}\,
\sigma^{y}_{j}
+
\Theta(t)\,f(t)\,
\sum_{\langle i j k \rangle}\,
\sigma^{z}_{i}
\,
\sigma^{z}_{j}
\,
\sigma^{z}_{k}
\,,
\end{split}•
\end{equation}
where the summation in the second term runs over
all the triangles of the lattice and we assume the same form
for $f(t)$ as in Eq.~\eqref{eq: 1D Hamiltonian in the lab frame}.
Interestingly, for
$
\lambda \approx 0.51
$
and
$
\varphi \approx \pm 0.27 \pi
$,
we find that (see Appendix B)
\begin{equation}
\label{eq: Floquet SPT Hamiltonian in 2D}
\begin{split}
\mathcal{H}^{(0)}_{\rm{F}}
\approx
h_{\rm{eff}}
\,
\sum_{j}\,
\sigma^{x}_{j}\,
e^
{
i\,\frac{\pi}{4}\,
\sum_{\langle \ell \ell' \rangle; j}\,
\left(
1
-
\sigma^{z}_{\ell}\,\sigma^{z}_{\ell'}
\right)
}
\,.
\end{split}
\end{equation}
Remarkably, the Hamiltonian~(\ref{eq: Floquet SPT Hamiltonian in 2D}),
which involves up to seven-spin interactions, 
is the exactly-solvable model of a $\mathbb{Z}_{2}$ SPT paramagnet studied 
by Levin and Gu in Ref.~\cite{Levin-2012}.
The model~(\ref{eq: Floquet SPT Hamiltonian in 2D})
can be obtained from a trivial paramagnet by a product of local unitary transformations that each depend 
on three $\sigma^{z}$ spins (see Appendix B),
which then justifies the need for a three-spin interaction 
in~(\ref{eq: 2D driven Hamiltonian}).

We close by commenting on possible experimental realizations of SSPT phases.  Recent developments in quantum simulation with trapped ions~\cite{Kim2009,Edwards2010,Kim2010,lanyon} and superconducting quantum circuits~\cite{Roushan2014,Chen2014} have shown that it is possible to engineer tunable multi-spin interactions and transverse fields in a laboratory setting.  These developments suggest the possibility that SSPT phases like the ones discussed in this paper could be realized in an experiment, if the appropriate sinusoidal drive can be implemented.  The superconducting quantum circuit architecture described in Ref.~\cite{Chen2014} appears particularly well-suited to these purposes, as it was demonstrated in that work that the couplings between the superconducting qubits in that system can be tuned dynamically.  The periodic modulation of the interaction strength required by our proposal is already feasible in that setup, making it an ideal candidate for a possible experimental realization.  A thorough assessment of the suitability of this proposal for the experimental platforms mentioned above is necessary, but beyond the scope of this work.

To summarize, we have shown in this work that, by adding appropriately chosen periodically-driven multispin interactions to a trivial paramagnetic Hamiltonian, it is possible to realize SPT phases in the high-frequency limit.  We further illustrated via a 1D example that the SPT phase can be probed with stroboscopic measurements of local observables.  We also characterized numerically (and analytically, in the Appendices) the finite-frequency corrections to the pure infinite-frequency SSPT Hamiltonian, and found that driving frequencies of order a hundred times the characteristic bare energy scale of the problem are sufficient to observe signatures of the phase.  Finally, we illustrated how this construction can be extended to higher dimensions and different symmetry classes, such as the above example of the $\mathbb Z_2$ SPT phase in 2D.  In future work, it would be interesting to further explore the possibility of generating nontrivial patterns of entanglement by driving.  The work presented here can also be used as a springboard to future progress in the development of out-of-equilibrium and non-eigenstate probes of topological and symmetry-protected topological order.  In particular, it would be interesting to determine, along the lines of Ref.~\cite{vishwanath}, how the presence of disorder can protect SPT order in the novel context, explored here, of periodically driven systems.

\begin{acknowledgments}
We thank Marin Bukov and Anushya Chandran for useful discussions.  T.I. was supported by the National Science Foundation Graduate Research Fellowship
Program under Grant No. DGE-1247312, and C.C. was supported by DOE Grant DEF-06ER46316.  Research at the Perimeter Institute
is supported by the Government of Canada through Industry Canada and by the
Province of Ontario through the Ministry of Economic Development and Innovation.
\end{acknowledgments}

\appendix

\begin{widetext}

\section{One-Dimensional SSPT Hamiltonian}

\setcounter{equation}{0}
\makeatletter
\renewcommand{\theequation}{A\arabic{equation}}

\subsection{Stroboscopic Hamiltonian}
We derive an effective Hamiltonian that encapsulates the stroboscopic dynamics generated by
\begin{subequations}
\begin{equation}
\label{eq: 1D Hamiltonian in the lab frame - appendix}
\begin{split}
&\,
H_{1\rm{D}}(t)
=
h\,
\sum^{N}_{i=1}\,
\sigma^{x}_{i}
+
\Theta(t)\,f(t)\,
\sum^{N-1}_{i=1}\,
\sigma^z_{i}\,
\sigma^z_{i+1}
\,,
\end{split}
\end{equation}
with
\begin{equation}
f(t)
=
\lambda\,\omega\,
\cos(\omega t + \varphi)
\,,
\quad
\lambda > 0
\,.
\end{equation}
\end{subequations}•

To do this, we employ the time-dependent unitary transformation
\begin{equation}
U^{\,}_{\rm R}(t)
=
\exp
{
\Big[
\mathrm{i}\,
\int_{-\infty}^{t}
\mathrm d\pri t\, \Theta(t)\,f(t)
\,
\sum^{N-1}_{i=1}\sigma^z_{i}\sigma^z_{i+1}
\Big]
}
=
\exp
\Big[
\mathrm i\,g(t)\,\sum^{N-1}_{i=1}\sigma^z_{i}\sigma^z_{i+1}\Big]
\,,
\end{equation}
where
$
g(t)
=
\lambda\,
\left[
\sin
{
\left(
\omega\,t
+
\varphi
\right)
}
-
\sin
{
\varphi
}
\right]
$,
which transforms the Hamiltonian to the rotating frame as follows:
\begin{equation}
H_{\rm R}(t)
=
U^{\,}_{\rm R}(t)\, H_{1\rm{D}}(t)\, U^\dagger_{\rm R}(t)-\mathrm i\, U^{\,}_{\rm R}(t)\, \partial_t\, U^{\dagger}_{\rm R}(t)\\
=
h\, U^{\,}_{\rm R}(t)\(\sum^{N}_{i=1}\sigma^x_{i}\)U^\dagger_{\rm R}(t).
\end{equation}

Explicitly we find
\begin{equation}
\label{eq: 1D Hamiltonian in the rotating frame - appendix}
\begin{split}
H_{\rm R}(t) 
&\,
= 
\sum^{N-1}_{i=2}
\,
\Big\{
\cos^2(2g(t))
\,
\sigma^x_i
-
\sin^2(2g(t))
\,
\sigma^z_{i-1}\sigma^x_i\sigma^z_{i+1}
-
\cos(2g(t))\sin(2g(t))
\,
(\sigma^y_{i}\sigma^z_{i+1}+\sigma^z_{i-1}\sigma^y_{i})
\Big\}
\\
&\,
\quad\quad\quad
+
\cos(2g(t))
\,
\left(
\sigma^{x}_{1}
+
\sigma^{x}_{N}
\right)
-
\sin(2g(t))
\,
\left(
\sigma^{y}_{1}
\,
\sigma^{z}_{2}
+
\sigma^{z}_{N-1}
\,
\sigma^{y}_{N}
\right)
\,.
\end{split}
\end{equation}
The time average of
Eq.~(\ref{eq: 1D Hamiltonian in the rotating frame - appendix})
yields
\begin{equation}
\begin{split}
\label{zeroth order magnus - appendix}
\mathcal{H}^{(0)}_{\rm F} 
&\,
= 
\sum^{N-1}_{i=2}\,
\Big\{
\,
a(\lambda,\varphi)\,
\sigma^{x}_{i}
-
b(\lambda,\varphi)
\,
\sigma^z_{i-1}\sigma^x_i\sigma^z_{i+1}
-
c(\lambda,\varphi)
\,
\left(
\,
\sigma^y_{i}\sigma^{z}_{i+1}+\sigma^{z}_{i-1}\sigma^y_{i}
\,
\right)
\,
\Big\}
\\
&\,
+
d(\lambda,\varphi)
\,
\left(
\sigma^{x}_{1}
+
\sigma^{x}_{N}
\right)
-
e(\lambda,\varphi)
\,
\left(
\sigma^{y}_{1}
\,
\sigma^{z}_{2}
+
\sigma^{z}_{N-1}
\,
\sigma^{y}_{N}
\right)
\,,
\end{split}•
\end{equation}
where the coefficients 
$
a(\lambda,\varphi), ... , e(\lambda,\varphi)
$
are given by
\begin{subequations}
\label{coefficients}
\begin{equation}
a(\lambda,\varphi) 
=
1
-
b(\lambda,\varphi)
= 
\frac{1}{2\pi}\,
\int_{0}^{2\pi}
\mathrm d\tau\, 
\cos^{2}
{
\big[
2\lambda
G_{\varphi}(\tau)
\big]
}
\,,
\end{equation}
\begin{equation}
c(\lambda,\varphi) 
= 
\frac{1}{2\pi}\,
\int_{0}^{2\pi}
\mathrm d\tau\, 
\frac{1}{2}
\sin
{
\big[
4\lambda
G_{\varphi}(\tau)
\big]
}
\,,
\end{equation}
\begin{equation}
d(\lambda,\varphi) 
= 
\frac{1}{2\pi}\,
\int_{0}^{2\pi}
\mathrm d\tau\, 
\cos
{
\big[
2\lambda
G_{\varphi}(\tau)
\big]
}
\,,
\end{equation}
\begin{equation}
e(\lambda,\varphi) 
= 
\frac{1}{2\pi}\,
\int_{0}^{2\pi}
\mathrm d\tau\, 
\sin
{
\big[
2\lambda
G_{\varphi}(\tau)
\big]
}
\,,
\end{equation}
\end{subequations}
where
$
G_{\varphi}(\tau)
=
\sin
{
\left(
\tau
+
\varphi
\right)
}
-
\sin
{
\varphi
}
$. 
Observe that for the choice $\varphi=0$, the two-body terms that break the $\mathbb Z_2\times\mathbb Z_2$ symmetry vanish and we recover the stroboscopic Hamiltonian in Eq.~\eqref{eq: finite chain} of the main text.

\subsection{Leading Finite-Frequency Correction to the SSPT Hamiltonian}

\begin{figure}
\centering
\includegraphics[width=.6\textwidth]{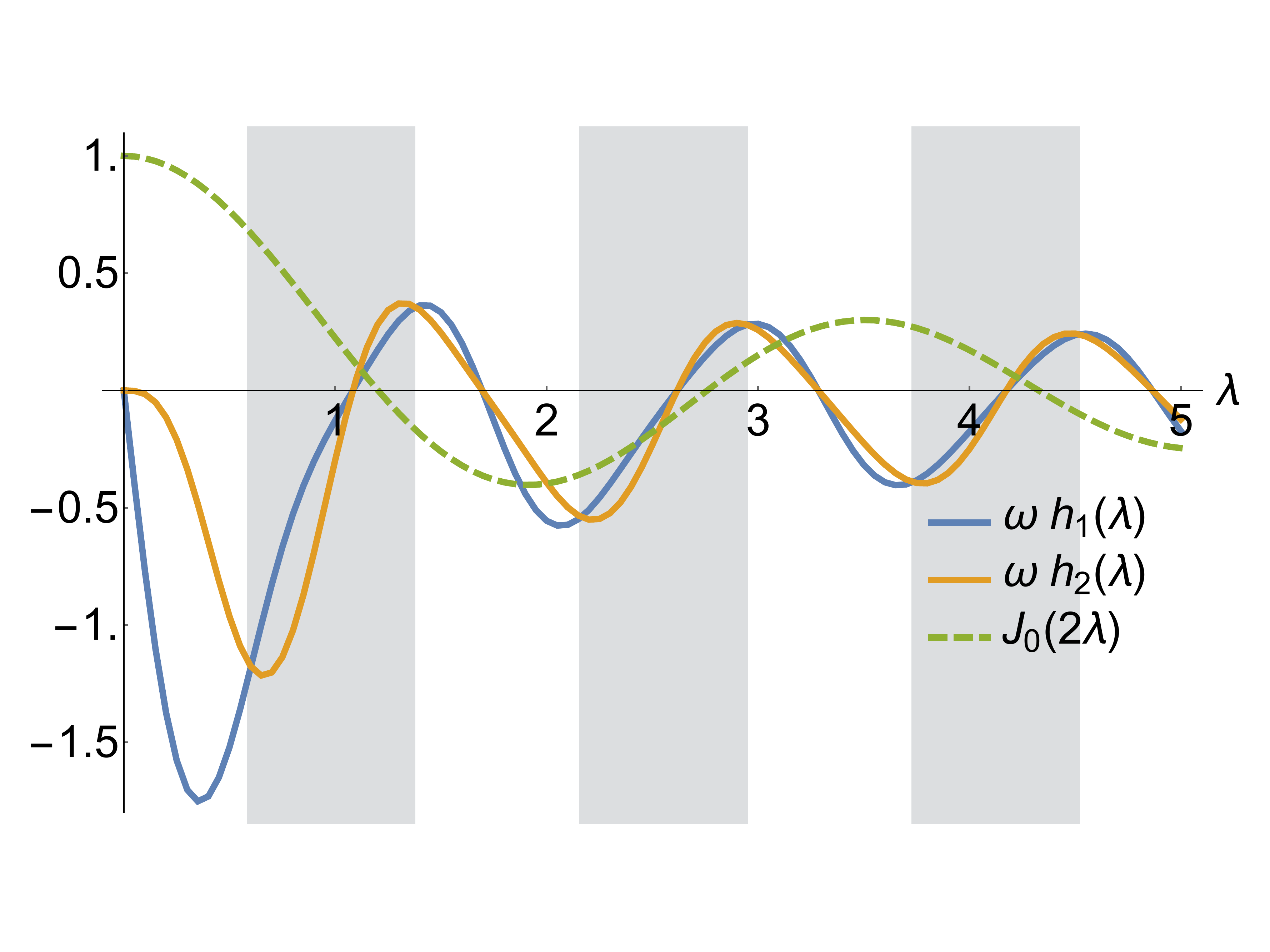}
\caption{(Color online) Couplings in the first-order correction to the Magnus expansion, \textit{c.f.}~Eq.~\eqref{first-order magnus}, as functions of the scaled driving amplitude $\lambda$. White and gray regions are used to distinguish the trivial and stroboscopic SPT phases, as in Fig.~\ref{fig: couplings}, and the Bessel function $J_0(2\lambda)$ is plotted for reference.}\label{fig: first-order couplings}
\end{figure}

We now present the order-$1/\omega$ Magnus correction to the SSPT Hamiltonian, namely
\begin{align}
\mathcal{H}^{(1)}_{\rm F}&=-\mathrm i\, \frac{1}{2\, T}\int_{0}^{T}\mathrm d t_1\int_{0}^{t_1}\mathrm dt_2\, [H_{\rm R}(t_1),H_{\rm R}(t_2)],
\end{align}
where $H_{\rm R}(t)$ is given by Eq.~\eqref{eq: H(t) in the rotating frame}.  We will work exclusively with an infinite chain in this section, as our aim is only to show that the bulk $\mathbb Z_2\times\mathbb Z_2$ symmetry is broken by this correction.  After calculating the necessary commutators, we find that
\begin{subequations}\label{first-order magnus}
\begin{align}
\mathcal{H}^{(1)}_{\rm F}&=h^{\,}_{1}\sum_i\, (\sigma^z_{i}\sigma^z_{i+1}-\sigma^y_i\sigma^y_{i+1})+ h^{\,}_2\sum_i\, (\sigma^z_{i-1}\sigma^x_i\sigma^x_{i+1}\sigma^z_{i+2} -\sigma^z_{i-1}\sigma^z_{i}),
\end{align}
where the coefficients are given by
\begin{align}
h^{\,}_1&=-\frac{1}{\pi\omega}\int_0^{2\pi}\mathrm d\tau_1\int_0^{\tau_1}\mathrm d\tau_2 \, \cos(2\lambda\sin\tau_1)\cos(2\lambda\sin\tau_2)\sin[2\lambda(\sin\tau_2-\sin\tau_1)]\\
h^{\,}_2 &=\frac{1}{\pi\omega}\int_0^{2\pi}\mathrm d\tau_1\int_0^{\tau_1}\mathrm d\tau_2 \, \sin(2\lambda\sin\tau_1)\sin(2\lambda\sin\tau_2)\sin[2\lambda(\sin\tau_2-\sin\tau_1)].
\end{align}
\end{subequations}
These coefficients are plotted as functions of $\lambda$ in Fig.~\ref{fig: first-order couplings}.  Observe that each term above breaks the $\mathbb{Z}_2\times\mathbb{Z}_2$ symmetry.  The $\mathbb{Z}_2\times\mathbb{Z}_2$ symmetry of the zeroth-order Hamiltonian \eqref{eq: finite chain} is therefore an emergent symmetry that appears only at high frequencies.

\section{Two-Dimensional SSPT Hamiltonian}

\setcounter{equation}{0}
\makeatletter
\renewcommand{\theequation}{B\arabic{equation}}

\subsection{Exactly Solvable $\mathbb{Z}_{2}$ SPT Model}

In this section, we review the 2D $\mathbb Z_2$ SPT model introduced by Levin and Gu in Ref.~\cite{Levin-2012}.  We start with the trivial paramagnetic Hamiltonian on the triangular lattice (see Fig.~\ref{fig: triangular lattice}),
\begin{equation}
\label{eq: Trivial paramagnet in 2D - appendix}
\begin{split}
H_{0}
=
-
\sum_{j}\,\sigma^{x}_{j}
\,,
\end{split}•
\end{equation}•
and the 2D $\mathbb{Z}_{2}$ SPT Hamiltonian~\cite{Levin-2012}
\begin{equation}
\label{eq: SPT paramagnet in 2D - appendix}
\begin{split}
H_{2D,SPT}
=
-
\sum_{j}\,\mathcal{O}_{j}
=
\sum_{j}\,
\sigma^{x}_{j}\,
e^
{
\mathrm i\,\frac{\pi}{4}\,
\sum_{\langle \ell \ell' \rangle; j}\,
\left(
1
-
\sigma^{z}_{\ell}\,\sigma^{z}_{\ell'}
\right)
}
\,,
\end{split}
\end{equation}
where the sum over $\ell, \ell'$ 
in Eq.~(\ref{eq: SPT paramagnet in 2D - appendix})
extends over pairs of nearest neighbor spins
around the spin at site $j$ as depicted in Fig.~\ref{fig: triangular lattice}.
The Hamiltonian 
Eq.~(\ref{eq: SPT paramagnet in 2D - appendix})
is invariant under spin flips generated by 
$
S_{\mathbb{Z}_{2}}
=
\prod_{j}\,
\sigma^{x}_{j}
$.
\begin{figure}
\centering
\includegraphics[width=.4\textwidth]{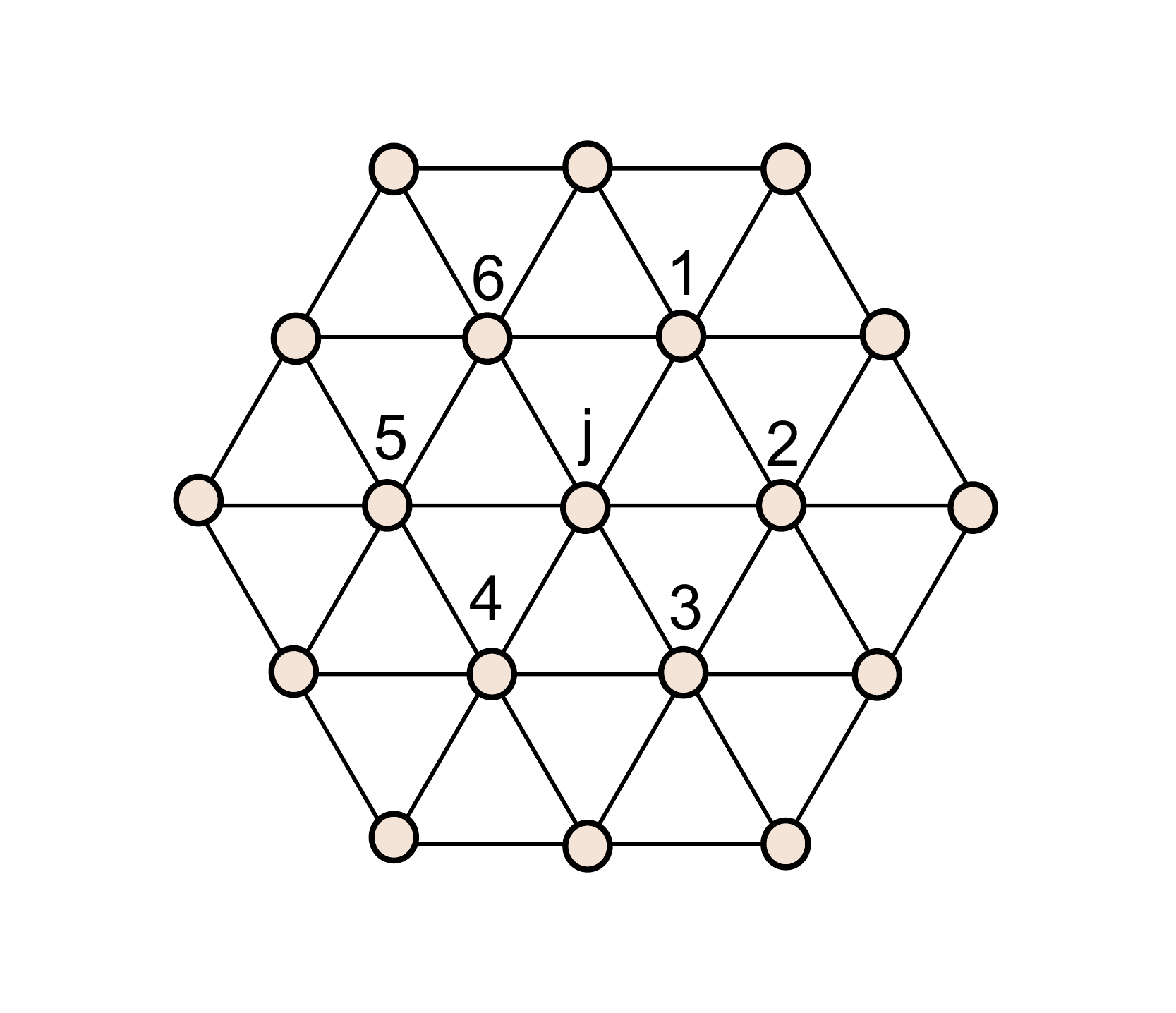}
\caption
{
Triangular lattice where the model
Eq.~(\ref{eq: SPT paramagnet in 2D - appendix})
and the driven Hamiltonian
Eq.~(\ref{eq: 2D driven Hamiltonian - appendix})
are defined.
The sites nearest neighbors to site $j$ are labeled $1$ through $6$.
}
\label{fig: triangular lattice}
\end{figure}

The Hamiltonians 
Eq.~(\ref{eq: Trivial paramagnet in 2D - appendix})
and~(\ref{eq: SPT paramagnet in 2D - appendix})
are related by the unitary transformation 
\begin{equation}
\label{eq: Unitary transformation to the 2D SPT paramagnet - appendix}
\mathbb{W}
=
\prod_{j}\,
e^
{
-\mathrm i\,
\frac{\pi}{2}
\left(
1
-
\sigma^{z}_{j}
\right)
-\mathrm i
\,\frac{\pi}{8}\,
\sigma^{z}_{j}\,
\sum_{\langle \ell \ell' \rangle; j}\,
\left(
1
-
\sigma^{z}_{\ell}\,\sigma^{z}_{\ell'}
\right)
}
\,,
\end{equation}•
that implements
\begin{equation}
\begin{split}
\mathcal{O}_{j}
=
\mathbb{W}
\,
\sigma^{x}_{j}
\,
\mathbb{W}^{-1}
=
-
\sigma^{x}_{j}
\,
e^
{
\mathrm i\,\frac{\pi}{4}\,
\sum_{\langle \ell \ell' \rangle; j}\,
\left(
1
-
\sigma^{z}_{\ell}\,\sigma^{z}_{\ell'}
\right)
}
\,.
\end{split}•
\end{equation}•

Now for every site $j$ we expand the exponent:
\begin{equation}
\begin{split}
&\,
e^
{
\mathrm i\,\frac{\pi}{4}\,
\sum_{\langle \ell \ell' \rangle; j}\,
\left(
1
-
\sigma^{z}_{\ell}\,\sigma^{z}_{\ell'}
\right)
}
=
e^
{
\mathrm i\,\frac{3\pi}{2}
}
\,
\prod_{\langle \ell \ell' \rangle; j}
\,
\left[
\cos{(\pi/4)}
-
\mathrm i\,\sin{(\pi/4)}\,\sigma^{z}_{\ell}\,\sigma^{z}_{\ell'}
\right]
\\
&\,=
\frac{1}{4}
\Big\{
\sigma^{z}_{1}\,\sigma^{z}_{2}\,\sigma^{z}_{3}\,\sigma^{z}_{4}\,\sigma^{z}_{5}\,\sigma^{z}_{6}
+
\sigma^{z}_{1}\,\sigma^{z}_{2}\,\sigma^{z}_{3}\,\sigma^{z}_{5}
+
\sigma^{z}_{1}\,\sigma^{z}_{3}\,\sigma^{z}_{4}\,\sigma^{z}_{5}
+
\sigma^{z}_{1}\,\sigma^{z}_{2}\,\sigma^{z}_{4}\,\sigma^{z}_{6}
+
\sigma^{z}_{2}\,\sigma^{z}_{3}\,\sigma^{z}_{4}\,\sigma^{z}_{6}
+
\sigma^{z}_{1}\,\sigma^{z}_{3}\,\sigma^{z}_{5}\,\sigma^{z}_{6}
+
\sigma^{z}_{2}\,\sigma^{z}_{4}\,\sigma^{z}_{5}\,\sigma^{z}_{6}
\\
&\,
\quad\quad
+
\sigma^{z}_{1}\,\sigma^{z}_{4}
+
\sigma^{z}_{2}\,\sigma^{z}_{5}
+
\sigma^{z}_{3}\,\sigma^{z}_{6}
-
\left(
\sigma^{z}_{1}\,\sigma^{z}_{2}
+
\sigma^{z}_{2}\,\sigma^{z}_{3}
+
\sigma^{z}_{3}\,\sigma^{z}_{4}
+
\sigma^{z}_{4}\,\sigma^{z}_{5}
+
\sigma^{z}_{5}\,\sigma^{z}_{6}
+
\sigma^{z}_{6}\,\sigma^{z}_{1}
\right)
\Big\}
\,,
\end{split}•
\end{equation}•
where
$
\sigma^{z}_{1},
...,
\sigma^{z}_{6}
$
denote the six spin operators around the site $j$, as in 
Fig.~\ref{fig: triangular lattice}.

\subsection{Driven Three-Spin Interaction}

Motivated by the unitary transformation 
Eq.~(\ref{eq: Unitary transformation to the 2D SPT paramagnet - appendix}),
we are led to consider a time dependent three-spin interaction
\begin{subequations}
\begin{equation}
\label{eq: 2D driven Hamiltonian - appendix} 
\begin{split}
H_{2D}(t)
=
-
h\,
\sum_{j}\,
\sigma^{y}_{j}
+
\Theta(t)\,f(t)\,
\sum_{\langle i j k \rangle}\,
\sigma^{z}_{i}
\,
\sigma^{z}_{j}
\,
\sigma^{z}_{k}
\,,
\end{split}•
\end{equation}•
where the summation
in the second term 
runs over every triangle of the lattice
and
\begin{equation}
f(t)
=
\lambda\,\omega\,
\cos(\omega t + \varphi)
\,,
\quad
\lambda > 0
\,.
\end{equation}
\end{subequations}•

The unitary transformation to the rotating frame
\begin{equation}
U^{\,}_{\rm R}(t)
=
\exp
\Big[
\mathrm i\, 
g(t)
\,
\sum_{\langle i j k \rangle}\,
\sigma^{z}_{i}
\,
\sigma^{z}_{j}
\,
\sigma^{z}_{k}
\Big]
\,,
\end{equation}
where
$
g(t)
=
\lambda\,
\left[
\sin
{
\left(
\omega\,t
+
\varphi
\right)
}
-
\sin
{
\varphi
}
\right]
$,
yields the rotating-frame Hamiltonian
\begin{equation}
\label{eq: 2D Hamiltonian in the rotating frame - appendix}
\begin{split}
H_{\rm{R}}(t)
=
U_{\rm{R}}(t)
\,
\left(
-
h\,
\sum_{j}
\,
\sigma^{y}_{j}
\right)
\,
U^{\dagger}_{\rm{R}}(t)
\,.
\end{split}•
\end{equation}•
The relevant object to compute is then 
\begin{equation}
\begin{split}
U_{R}(t)
\,
\sigma^{y}_{j}
\,
U_{R}^{\dagger}(t)
&\,
=
\sigma^{y}_{j}
\,
\exp
\left[
-\mathrm i\, 2\,g(t)\,\sigma^{z}_{j}\,\sum_{\langle\ell \ell' j \rangle}\,\sigma^{z}_{\ell}\,\sigma^{z}_{\ell'}
\right]
\\
&\,=
\sigma^{y}_{j}\,
\prod_{\langle \ell \ell' j \rangle}
\,
\left[
\cos(2 g(t))
-
\mathrm i\,\sin(2 g(t))\,
\sigma^{z}_{j}\,\sigma^{z}_{\ell}\,\sigma^{z}_{\ell'}
\right]
\\
&\,
\equiv
\sigma^{y}_{j}\,\mathcal{A}_{j}(t)
\,.
\end{split}•
\end{equation}•

Explicitly, we have
\begin{subequations}
\label{eq: A operator 2D rotating frame - appendix}
\begin{equation}
\begin{split}
\mathcal{A}_{j}(t)
=
\mathcal{A}_{j}^{\rm{I}}(t)
+
\mathcal{A}_{j}^{\rm{II}}(t)
\,,
\end{split}•
\end{equation}•
where
\begin{equation}
\begin{split}
\mathcal{A}_{j}^{\rm{I}}(t)
&\,=
\mathrm i\,\sigma^{z}_{j}\,
\Big[
\beta_{1}(t)
\,
\Big(
\sigma^{z}_{1}\,\sigma^{z}_{4}
+
\sigma^{z}_{2}\,\sigma^{z}_{5}
+
\sigma^{z}_{3}\,\sigma^{z}_{6}
\\
&\,\quad\quad\quad\quad\quad\quad
+
\sigma^{z}_{1}\,\sigma^{z}_{2}\,\sigma^{z}_{3}\,\sigma^{z}_{5}
+
\sigma^{z}_{1}\,\sigma^{z}_{3}\,\sigma^{z}_{4}\,\sigma^{z}_{5}
+
\sigma^{z}_{1}\,\sigma^{z}_{2}\,\sigma^{z}_{4}\,\sigma^{z}_{6}
\\
&\,\quad\quad\quad\quad\quad\quad
+
\sigma^{z}_{2}\,\sigma^{z}_{3}\,\sigma^{z}_{4}\,\sigma^{z}_{6}
+
\sigma^{z}_{1}\,\sigma^{z}_{3}\,\sigma^{z}_{5}\,\sigma^{z}_{6}
+
\sigma^{z}_{2}\,\sigma^{z}_{4}\,\sigma^{z}_{5}\,\sigma^{z}_{6}
\\
&\,\quad\quad\quad\quad\quad\quad
+
\sigma^{z}_{1}\,\sigma^{z}_{2}\,\sigma^{z}_{3}\,\sigma^{z}_{4}\,\sigma^{z}_{5}\,\sigma^{z}_{6}
\Big)
\\
&\,\quad\quad\quad
-\beta_{2}(t)
\,
\left(
\sigma^{z}_{1}\,\sigma^{z}_{2}
+
\sigma^{z}_{2}\,\sigma^{z}_{3}
+
\sigma^{z}_{3}\,\sigma^{z}_{4}
+
\sigma^{z}_{4}\,\sigma^{z}_{5}
+
\sigma^{z}_{5}\,\sigma^{z}_{6}
+
\sigma^{z}_{6}\,\sigma^{z}_{1}
\right)
\Big]
\,,
\end{split}•
\end{equation}•
\begin{equation}
\begin{split}
\mathcal{A}_{j}^{\rm{II}}(t)
&\,=
\beta_{3}(t)
+
\beta_{4}(t)
\,
\Big(
\sigma^{z}_{1}\, \sigma^{z}_{2} \, \sigma^{z}_{3} \, \sigma^{z}_{4}
+
\sigma^{z}_{1} \, \sigma^{z}_{2}\, \sigma^{z}_{3} \, \sigma^{z}_{6}
+
\sigma^{z}_{1}\,\sigma^{z}_{2}\,\sigma^{z}_{4}\,\sigma^{z}_{5}
\\
&\,\quad\quad\quad\quad\quad\quad\quad
+
\sigma^{z}_{1}\,\sigma^{z}_{2}\,\sigma^{z}_{5}\,\sigma^{z}_{6}
+
\sigma^{z}_{1}\,\sigma^{z}_{3}\,\sigma^{z}_{4}\,\sigma^{z}_{6}
+
\sigma^{z}_{1}\,\sigma^{z}_{4}\,\sigma^{z}_{5}\,\sigma^{z}_{6}
\\
&\,\quad\quad\quad\quad\quad\quad\quad
+
\sigma^{z}_{2}\,\sigma^{z}_{3}\,\sigma^{z}_{4}\,\sigma^{z}_{5}
+
\sigma^{z}_{2}\,\sigma^{z}_{3}\,\sigma^{z}_{5}\,\sigma^{z}_{6}
+
\sigma^{z}_{3}\,\sigma^{z}_{4}\,\sigma^{z}_{5}\,\sigma^{z}_{6}
\\
&\,\quad\quad\quad\quad\quad\quad\quad
+
\sigma^{z}_{1}\,\sigma^{z}_{3}
+
\sigma^{z}_{1}\,\sigma^{z}_{5}
+
\sigma^{z}_{2}\,\sigma^{z}_{4}
+
\sigma^{z}_{2}\,\sigma^{z}_{6}
+
\sigma^{z}_{3}\,\sigma^{z}_{5}
+
\sigma^{z}_{4}\,\sigma^{z}_{6}
\Big)
\,,
\end{split}•
\end{equation}•
where
\begin{equation}
\begin{split}
&\,
\beta_{1}(t)
=
2\,c^{3}(t)\,s^{3}(t)
\,,
\\
&\,
\beta_{2}(t)
=
c^{}(t)\,s^{5}(t)
+
c^{5}(t)\,s^{}(t)
\,,
\\
&\,
\beta_{3}(t)
=
c^{6}(t) - s^{6}(t)
\,,
\\
&\,
\beta_{4}(t)
=
c^{2}(t)\,s^{4}(t)
-
c^{4}(t)\,s^{2}(t)
\,,
\end{split}•
\end{equation}•
\end{subequations}•
and we use the shorthand notation
$
c(t)
\equiv
\cos(2 g(t))
$
and
$
s(t)
\equiv
\sin(2 g(t))
$.

The Floquet Hamiltonian at infinite frequency, obtained
from the time average of the Hamiltonian 
Eq.~(\ref{eq: 2D Hamiltonian in the rotating frame - appendix}),
\begin{equation}
\label{eq: Floquet Hamiltonian 2D - appendix}
\begin{split}
\mathcal{H}^{(0)}_{\rm{F}}
&\,
=
-h\,
\sum_{j}\,
\sigma^{y}_{j}\,
\Big(
\frac{1}{T}\,
\int^{T}_{0}\,
\mathrm{d} t\,
\mathcal{A}_{j}(t)
\Big)
\,,
\end{split}•
\end{equation}•
upon using Eq.~(\ref{eq: A operator 2D rotating frame - appendix}),
depends on the following parameters
\begin{subequations}
\begin{equation}
\begin{split}
\beta_{1}(\lambda,\varphi)
&\,=
\frac{1}{2\,\pi}\,
\int^{2\,\pi}_{0}\, d \tau \,
2\,
\cos^{3}
{
\Big[
2 \lambda G_{\varphi}(\tau)
\Big]
}
\,
\sin^{3}
{
\Big[
2 \lambda G_{\varphi}(\tau)
\Big]
}
\,,
\end{split}•
\end{equation}•
\begin{equation}
\begin{split}
\beta_{2}(\lambda,\varphi)
&\,=
\frac{1}{2\,\pi}\,
\int^{2\,\pi}_{0}\, d \tau \,
\Big\{
\cos^{}
{
\Big[
2 \lambda G_{\varphi}(\tau)
\Big]
}
\,
\sin^{5}
{
\Big[
2 \lambda G_{\varphi}(\tau)
\Big]
}
+
\cos^{5}
{
\Big[
2 \lambda G_{\varphi}(\tau)
\Big]
}
\,
\sin^{}
{
\Big[
2 \lambda G_{\varphi}(\tau)
\Big]
}
\Big\}
\,,
\end{split}•
\end{equation}•
\begin{equation}
\beta_{3}(\lambda,\varphi)
=
\frac{1}{2\,\pi}\,
\int^{2\,\pi}_{0}\, d \tau \,
\Big\{
\cos^{6}
{
\Big[
2 \lambda G_{\varphi}(\tau)
\Big]
}
-
\sin^{6}
{
\Big[
2 \lambda G_{\varphi}(\tau)
\Big]
}
\Big\}
\,,
\end{equation}•
\begin{equation}
\begin{split}
\beta_{4}(\lambda,\varphi)
&\,=
\frac{1}{2\,\pi}\,
\int^{2\,\pi}_{0}\, d \tau \,
\Big\{
\cos^{2}
{
\Big[
2 \lambda G_{\varphi}(\tau)
\Big]
}
\,
\sin^{4}
{
\Big[
2 \lambda G_{\varphi}(\tau)
\Big]
}
-
\cos^{4}
{
\Big[
2 \lambda G_{\varphi}(\tau)
\Big]
}
\,
\sin^{2}
{
\Big[
2 \lambda G_{\varphi}(\tau)
\Big]
}
\Big\}
\,,
\end{split}•
\end{equation}•
\end{subequations}•
where
$
G_{\varphi}(\tau)
=
\sin
{
\left(
\tau
+
\varphi
\right)
}
-
\sin
{
\varphi
}
$.
Whenever
\begin{subequations}
\label{eq: condition on betas - appendix}
\begin{equation}
\label{eq: condition beta 1 and beta 2}
\beta_{1}(\lambda^{*},\varphi^{*}) 
= 
\beta_{2}(\lambda^{*},\varphi^{*}) 
\equiv
\beta^{*}
\neq 0
\,,
\end{equation}
\begin{equation}
\label{eq: condition beta 3 and beta 4}
\beta_{3}(\lambda^{*},\varphi^{*})
=
\beta_{4}(\lambda^{*},\varphi^{*}) = 0
\,,
\end{equation}•
\end{subequations}
the Hamiltonian 
Eq.~(\ref{eq: Floquet Hamiltonian 2D - appendix})
acquires the form
\begin{equation}
\begin{split}
\mathcal{H}^{(0)}_{F}
=
4\,\beta^{*}\,h\,
\sum_{j}\,
\sigma^{x}_{j}\,
e^
{
\mathrm i\,\frac{\pi}{4}\,
\sum_{\langle \ell \ell' \rangle; j}\,
\left(
1
-
\sigma^{z}_{\ell}\,\sigma^{z}_{\ell'}
\right)
}
\,,
\end{split}
\end{equation}
which is the same model Eq.~(\ref{eq: SPT paramagnet in 2D - appendix})
shown in Ref.~\cite{Levin-2012} to describe the 2D SPT paramagnet with $\mathbb{Z}_{2}$ symmetry.
We have found numerically that condition
Eq.~(\ref{eq: condition on betas - appendix})
is satisfied, for example, for 
$
\lambda^{*}
\approx
0.51
$
and
$
\varphi
\approx
\pm 
0.27\,\pi
$.
It is fundamental to stress that even though the driven Hamiltonian Eq.~(\ref{eq: 2D driven Hamiltonian - appendix}) does not have the
$\mathbb{Z}_{2}$ symmetry, this symmetry emerges in the $\omega \rightarrow \infty$ limit.

\end{widetext}


\end{document}